\documentclass{article}%
\usepackage{amsfonts}
\usepackage{amsmath}
\usepackage{amssymb}
\usepackage{graphicx}%
\providecommand{\U}[1]{\protect\rule{.1in}{.1in}}
\newtheorem{theorem}{Theorem}

\newtheorem{corollary}[theorem]{Corollary}

\newtheorem{definition}[theorem]{Definition}

\newtheorem{lemma}[theorem]{Lemma}

\newenvironment{proof}[1][Proof]{\noindent\textbf{#1.} }{\ \rule{0.5em}{0.5em}}
\begin{document}

\title{From St\"{a}ckel systems to integrable hierarchies of PDE's:\\Benenti class of separation relations}
\author{Maciej B\l aszak\footnotemark\\Institute of Physics, A. Mickiewicz University\\Umultowska 85, 61-614 Pozna\'{n}, Poland
\and Krzysztof Marciniak\thanks{*Partially supported by KBN Research Project 1 P03B
111 27 and by Swedish Research Science Council grant no 2004-6920.}\\Department of Science and Technology \\Campus Norrk\"{o}ping, Link\"{o}ping University\\601-74 Norrk\"{o}ping, Sweden}
\maketitle

\begin{abstract}
We propose a general scheme of constructing\ of soliton hierarchies from
finite dimensional St\"{a}ckel systems and related separation relations. In
particular, we concentrate on the simplest class of separation relations,
called Benenti class, i.e. certain St\"{a}ckel systems with quadratic in
momenta integrals of motion.

\end{abstract}

\section{Introduction}

The theory of integrable nonlinear evolution equations has a long history as a
part of many branches of theoretical physics and applied mathematics.
Generally it can be divided in two parts: the theory of integrable nonlinear
ordinary differential equations (ODE's) and the theory of integrable nonlinear
partial differential equations (PDE's). Within the first class of equations
(ODE's) we will consider finite dimensional Hamiltonian systems, integrable by
the Hamilton-Jacobi method, called \emph{St\"{a}ckel systems}, while within
the second class (PDE's) we will consider (1+1)-dimensional field systems,
having infinite hierarchy of commuting symmetries and called further for
simplicity \emph{soliton systems}. The solvability by quadratures of some
class of finite dimensional systems by the Hamilton-Jacobi method, laid in the
19-th century one of the fundaments of analytical mechanics of integrable
systems, while the solvability by quadratures of some class of infinite
dimensional field systems by the Inverse Scattering Method, laid in second
half of the 20-th century one of the fundaments of the so called soliton
theory. \ 

During the last few decades both theories have been developed very intensively
using many common modern mathematical tools like Lax representation, r-matrix
theory, multi-Hamiltonian theory etc. In that time some links between both
theories were investigated. It was found (\cite{bogo}-\cite{stefmar}, see also
references in \cite{mb1}) that finite dimensional restrictions, invariant with
respect to the action of a given soliton system, like stationary flows,
restricted flows or constrained flows of Lax representation, are Liouville
integrable Hamiltonian systems of St\"{a}ckel type. Moreover, analytical
solutions of an appropriate finite dimensional systems are closely related to
a special class of solutions of related soliton systems, like for example
so-called finite-gap solutions \cite{its},\cite{zaharov}.

In the present paper we are interested in passing in the opposite direction -
building integrable hierarchies of PDE's from St\"{a}ckel systems \cite{a1}.
In that sense we would like to initiate a unified approach to St\"{a}ckel
ODE's and soliton PDE's. Our claim is the following: \emph{both a wide class
of St\"{a}ckel systems and a wide class of soliton systems can be constructed
form common fundamental objects known as\ separation relations} \emph{(or from
separation curves).}

The paper is organized as follows. Section 2 is devoted to general description
of the concept of separation relations. Section 3 explains the main ideas of
relating soliton systems with separation curves that are quadratic in momenta.
The idea is to apply to a set of Killing vector fields a set of invariants
generated by Euler-Lagrange equations associated with appropriately chosen
Lagrangian densities. This allows for elimination of some variables in our
Killing systems which leads to dispersive soliton hierarchies. Section 4 is a
brief introduction to what can be called Benenti class of St\"{a}ckel systems.
In section 5 we describe the structure of our systems in Vi\`{e}te
coordinates. In Section 6 we explain the details of our elimination procedure
which allows, in a systematic way, to construct soliton hierarchies. It is
divided into two subsections as the elimination procedure differs in case of
"positive" and "negative" (see below) separable potentials. Section 7
concludes the article with several examples.

\section{Separation relations}

Let us consider a $2n$-dimensional manifold $M$ equipped with a Poisson
operator $\Pi$ with some canonical (Darboux) coordinates labelled as $M\ni
u=(\mu,\lambda)$, with $\mu=(\mu_{1},\ldots,\mu_{n})$ and $\lambda
=(\lambda_{1},\ldots,\lambda_{n})$. The following definition introduces the
basic object of our considerations \cite{Sklyanin}:

\begin{definition}
\bigskip A set of $n$ relations of the form%
\begin{equation}
\varphi_{i}(\lambda_{i},\mu_{i},a_{1},\ldots,a_{n})=0\text{, }i=1,\ldots
,n\text{, \ }a_{i}\in\mathbf{R} \label{SRalg}%
\end{equation}
(each involving only one pair $\lambda_{i},\mu_{i}$ of canonical coordinates)
are called separation relations provided that the dependence of $\varphi$ on
$a$ is essential i.e. that $\det\left(  \frac{\partial\varphi_{i}}{\partial
a_{j}}\right)  \neq0$.
\end{definition}

The condition in (\ref{SRalg}) means that we can resolve the equations
(\ref{SRalg}) with respect to $a_{i}$ obtaining $a_{i}=H_{i}(\lambda,\mu)$,
$i=1,\ldots,n$. This defines some new functions $H_{i}(\lambda,\mu)$ that in
turn define the following Hamiltonian systems (evolutionary vector fields) on
$M$
\begin{equation}
u_{t_{i}}=\Pi\,dH_{i}=X_{H_{i}},\ \ \ \ \ \ \ i=1,...,n. \label{2.2}%
\end{equation}
If the functions $W_{i}(\lambda_{i},a)~$\ are solutions of a system of $n$
decuple \ ODE's obtained from (\ref{SRalg}) by substituting $\mu_{i}%
=\frac{\partial W_{i}(\lambda_{i},a)}{\partial\lambda_{i}}$%
\begin{equation}
\varphi_{i}\left(  \lambda_{i},\mu_{i}=\frac{\partial W_{i}(\lambda_{i}%
,a)}{\partial\lambda_{i}},a_{1},\ldots,a_{n}\right)  =0\text{, \ }i=1,...,n,
\label{SRdiff}%
\end{equation}
then the function $W(\lambda,a)=%
{\textstyle\sum\nolimits_{i=1}^{n}}
W_{i}(\lambda_{i},a)$ is an additively separable solution of \emph{all} the
equations (\ref{SRdiff}) and \emph{simultaneously} it is a solution of all
Hamilton-Jacobi equations
\begin{equation}
a_{i}=H_{i}\left(  \lambda,\mu=\frac{\partial W(\lambda,a)}{\partial\lambda
}\right)  \text{, \ \ }i=1,...,n \label{HJsim}%
\end{equation}
related with the Hamiltonians $H_{i}$ - simply because solving (\ref{SRalg})
to the form $a_{i}=H_{i}(\lambda,\mu)$ is a purely algebraic operation. Assume
now that $\det\left(  \frac{\partial^{2}W}{\partial\lambda_{i}\partial a_{j}%
}\right)  =\det\left(  \frac{\partial^{2}W_{i}}{\partial\lambda_{i}\partial
a_{j}}\right)  \neq0$. Then the Hamiltonians $H_{i}$ Poisson-commute since the
constructed function $W(\lambda,a)$ is a generating function for the canonical
transformation $(\lambda,\mu)\rightarrow(b,a)$ where
\begin{equation}
b_{i}=\frac{\partial W(\lambda,\alpha)}{\partial a_{i}}=t_{i}+const_{i}%
,\ \ \ \ i=1,...,n. \label{lin}%
\end{equation}
Equations (\ref{lin}) are implicit solutions of (\ref{2.2}) known as the
\emph{inversion Jacobi problem. }Thus, starting from a set of $n$ separation
relations we can create an $n$-dimensional separable Liouville system. All
systems separable in the sense of Hamilton-Jacobi theory can be obtained in
this way.

In an important case, when the functions $\varphi_{i}$ in (\ref{SRalg}) do not
depend on the index $i,$ the separation relations (\ref{SRalg}) can be
generated by taking $n$ copies of a curve in $\lambda$-$\mu$ plane:%
\begin{equation}
\varphi(\lambda,\mu,a_{1},\ldots,a_{n})=0\text{, \ }a_{i}\in\mathbf{R}
\label{SEPcurve}%
\end{equation}
called \emph{separation curve}.

Restricting our considerations to a subclass of (\ref{SRalg}), when all
separation relations are affine in $a_{i}=H_{i}$ with coefficients being
monomials in $\lambda$ and $\mu$, we obtain
\begin{equation}
\sum_{k=1}^{n}H_{k}\mu_{i}^{\alpha_{k}}\lambda_{i}^{\beta_{k}}=\psi
_{i}(\lambda_{i},\mu_{i}),\ \ i=1,...,n,\ \ \ \ \alpha_{k},\beta_{k}%
\in\mathbf{N} \label{2.7}%
\end{equation}
where $\psi_{i}$ are arbitrary smooth functions of two arguments. Equations
(\ref{2.7}) are called generalized \emph{St\"{a}ckel separation relations} and
the related dynamic systems, generated by Hamiltonian functions $H_{i},$ are
called the \emph{St\"{a}ckel separable} ones. To recover explicit St\"{a}ckel
form of Hamiltonians it is sufficient to solve the linear system (\ref{2.7})
with respect to $H_{i}$. If additionally $\psi_{i}(\lambda_{i},\mu_{i}%
)=\psi(\lambda_{i},\mu_{i})$ then the above separation conditions can be
represented by $n$ copies of the following separation curve:
\begin{equation}
\sum_{k=1}^{n}H_{k}\,\mu^{\alpha_{k}}\lambda^{\beta_{k}}=\psi(\lambda,\mu).
\label{2.8}%
\end{equation}
The separable systems that were most intensively studied in the last century
were one-particle dynamical systems on Riemannian manifolds with flat or
constant curvature metrics. All\ these systems can be obtained by choosing
$\alpha_{i}=0,\beta_{i}=n-i$, $i=1,...,n$ with $\psi$ quadratic in momenta
\begin{equation}
\psi(\lambda,\mu)=\frac{1}{2}f(\lambda)\mu^{2}+\gamma(\lambda). \label{2.8b}%
\end{equation}
This case will be considered in the next sections of this article.

We can now shortly present - by a simple example - the possibility of passing
from a separation curve to soliton systems \cite{a2}. Let us consider the
separation curve (\ref{2.8}) with $n=2,$ $\alpha_{1}=\alpha_{2}=0,$ $\beta
_{1}=1,$ $\beta_{2}=0$ and with $\psi$ in the form of (\ref{2.8b})%
\begin{equation}
H_{1}\lambda+H_{2}=\frac{1}{2}\lambda\mu^{2}+\lambda^{4}. \label{2.9}%
\end{equation}
The related separation conditions (\ref{2.7}) are%
\begin{equation}
\left\{
\begin{array}
[c]{c}%
H_{1}\lambda_{1}+H_{2}=\frac{1}{2}\lambda_{1}\mu_{1}^{2}+\lambda_{1}^{4}\\
H_{1}\lambda_{2}+H_{2}=\frac{1}{2}\lambda_{2}\mu_{2}^{2}+\lambda_{2}^{4}%
\end{array}
\right.  \label{2.10}%
\end{equation}
Solving this linear system with respect to $H_{1}$ and $H_{2}$ one gets the
Liouville integrable system (\ref{2.2}) on four dimensional phase space,
written in separation coordinates $(\lambda,\mu)$. The explicit form of
Hamiltonians $H_{i}$ is%
\[
H_{1}=\frac{1}{2}\frac{\lambda_{1}\mu_{1}^{2}-\lambda_{2}\mu_{2}^{2}%
+2\lambda_{1}^{4}-2\lambda_{2}^{4}}{\lambda_{1}-\lambda_{2}}\text{, \ \ }%
H_{2}=\frac{1}{2}\frac{\lambda_{1}\lambda_{2}\left(  \mu_{1}^{2}-\mu_{2}%
^{2}+2\lambda_{1}^{3}-2\lambda_{2}^{3}\right)  }{\lambda_{2}-\lambda_{1}}%
\]
The canonical transformation of the form
\[
q_{1}=\lambda_{1}+\lambda_{2},\ \ \ \frac{1}{4}q_{2}^{2}=-\lambda_{1}%
\lambda_{2},
\]%
\[
p_{1}=\frac{\lambda_{1}\mu_{1}}{\lambda_{1}-\lambda_{2}}+\frac{\lambda_{2}%
\mu_{2}}{\lambda_{2}-\lambda_{1}},\ \ \ \ p_{2}=\sqrt{-\lambda_{1}\lambda_{2}%
}\left(  \frac{\mu_{1}}{\lambda_{1}-\lambda_{2}}+\frac{\mu_{2}}{\lambda
_{2}-\lambda_{1}}\right)
\]
transforms the system to new coordinates $(q,p),$ with
\[
H_{1}=\frac{1}{2}p_{1}^{2}+\frac{1}{2}p_{2}^{2}+q_{1}^{3}+\frac{1}{2}%
q_{1}q_{2}^{2},\text{ \ \ }H_{2}=\frac{1}{2}q_{2}p_{1}p_{2}-\frac{1}{2}%
q_{1}p_{2}^{2}+\frac{1}{16}q_{2}^{4}+\frac{1}{4}q_{1}^{2}q_{2}^{2}.
\]
The function $H_{1}(q,p)$ turns out to be the Hamiltonian function of the
integrable case of the Henon-Heiles system, while $H_{2}(q,p)$ is the
additional involutive first integral of this system. Let us now denote the
evolution parameters $t_{1}$ and $t_{2}$ of the system by $x$ and $t$,
respectively. Then we obtain
\[
q_{1,x}=\frac{\partial H_{1}}{\partial p_{1}}=p_{1},\ \ \ \ q_{2,x}%
=\frac{\partial H_{1}}{\partial p_{2}}=p_{2},
\]%
\[
q_{1,t}=\frac{\partial H_{2}}{\partial p_{1}}=\frac{1}{2}q_{2}p_{2}%
,\ \ \ q_{2,t}=\frac{\partial H_{2}}{\partial p_{2}}=\frac{1}{2}q_{2}%
p_{1}-q_{1}p_{2},
\]
from which eliminating $p_{1}$ and $p_{2}$ we obtain a system of first order
PDE's for $q_{1}(x,t)$ and $q_{2}(x,t)$
\begin{equation}
q_{1,t}=\frac{1}{2}q_{2}q_{2},_{x}=\frac{1}{4}(q_{2}^{2})_{x},\ \ \ q_{2,t}%
=\frac{1}{2}q_{2}q_{1,x}-q_{1}q_{2,x}. \label{2.13}%
\end{equation}
Finally, we can eliminate $q_{2}$ through
\[
q_{1,xx}=p_{1,x}=-\frac{\partial H_{1}}{\partial q_{1}}=-3q_{1}^{2}-\frac
{1}{2}q_{2}^{2}\text{ }%
\]
\ which yields $q_{2}^{2}=-6q_{1}^{2}-2q_{1,xx}$ and then generate a higher
order (in $x-$derivatives) PDE. The first equation in (\ref{2.13}) turns then
into the famous KdV soliton system
\begin{equation}
q_{1,t}+\frac{1}{2}q_{1,xxx}+3q_{1}q_{1,x}=0, \label{2.14}%
\end{equation}
while the second equation in (\ref{2.13}) turns into a differential
consequence of the first one. Obviously, just from the presented construction,
there is no guarantee that equation (\ref{2.14}) is integrable. We can only
say that $q_{1}(x,t)$ calculated from the corresponding inverse Jacobi problem
is a nontrivial particular solution (one-gap solution) for the field system
(\ref{2.14}). To prove the integrability of (\ref{2.14}) one has to construct
some related infinite hierarchy of symmetries using some more regular procedure.

\section{From separation curves to constrained dispersionless systems}

In this paper we will concentrate on a special but important class of
separation curves with the function $\psi(\lambda,\mu)$ in (\ref{2.8}) being
quadratic in momenta $\mu$, (more precisely, of the form (\ref{2.8b})) and
with multipliers of Hamiltonian functions being monomials with respect to
$\lambda$
\begin{equation}
H_{1}\lambda^{\beta_{1}}+...+H_{n}\lambda^{\beta_{n}}=\frac{1}{2}\lambda
^{m}\mu^{2}+\lambda^{k}, \label{SC1}%
\end{equation}
where $\beta_{1}>\ldots>\beta_{n-1}>\beta_{n}=0$, $\beta_{i}\in\mathbf{N},$
$m,k\in\mathbf{Z}$ and $n\in\mathbf{N}$. Separable systems from this class
describe one-particle dynamics on Riemannian manifolds and belong to classical
St\"{a}ckel systems. Each class of these systems is labelled by a decreasing
sequence $(\beta_{1},...,\beta_{n})$ while members of a given class are
numbered by pairs $(m,k)\in\mathbf{Z}^{2}$. Taking $n$ copies of the curve
(\ref{SC1}) with variables $(\lambda,\mu)$ labelled within each copy as
$(\lambda_{i},\mu_{i})$, we obtain a system of $n$ separation relations in the
form of $n$ equations linear in the coefficients $H_{i}$. Solving it we obtain
$n$ functions $H_{r}^{(m,k)}=H_{r}^{(m,k)}(\lambda,\mu)$ of the form%
\begin{equation}
H_{r}^{(m,k)}=\frac{1}{2}\mu^{T}K_{r}G^{(m)}\mu+V_{r}^{(k)}\text{,
\ }r=1,\ldots,n\text{, \ }m,k\in\mathbf{Z} \label{Ham}%
\end{equation}
where we denote $\lambda=(\lambda_{1},\ldots,\lambda_{n})^{T}$ and $\mu
=(\mu_{1},\ldots,\mu_{n})^{T}$. The functions (\ref{Ham}) can be interpreted
as $n$ Hamiltonians on the phase space $T^{\ast}\mathcal{Q}$ cotangent to a
Riemannian manifold $\mathcal{Q}$ equipped with the contravariant metric
tensor $G^{(m)}$. These Hamiltonians are in involution with respect to the
canonical Poisson bracket on $T^{\ast}\mathcal{Q}$. Moreover, they are
separable in the sense of Hamilton-Jacobi theory since they by the very
definition satisfy St\"{a}ckel relations (\ref{SC1}). The objects $K_{r}$ in
(\ref{Ham}) can be interpreted as $(1,1)$-type Killing tensors on
$\mathcal{Q}$. The scalar functions $V_{r}^{(k)}$ are separable potentials.
Further, all the metric tensors $G^{(m)}$ and all the Killing tensors $K_{r}$
\ are diagonal in $\lambda$-variables so that:
\begin{equation}
K_{r}=\text{diag}(v_{r}^{1},\ldots,v_{r}^{n}) \label{Kdiag}%
\end{equation}
where $v_{r}^{i}$ are eigenvalues of $K_{r}$. We will constantly assume that
these eigenvalues are single.

The set (\ref{Ham}) of $n$ Hamiltonian functions leads to $n$ Hamiltonian
systems on $T^{\ast}\mathcal{Q}$ of the form%

\begin{equation}
\lambda_{t_{r}}=\frac{\partial H_{r}^{(k,m)}}{\partial\mu},\ \ \mu_{t_{r}%
}=-\frac{\partial H_{r}^{(k,m)}}{\partial\lambda},\ \ \ \ r=1,...,n.
\label{hamsys}%
\end{equation}
Let us now call the variable $t_{1}$ as $x$; $t_{1}\equiv x.$ Since all the
Hamiltonians $H_{r}^{(k,m)}$ (for fixed $k$ and $m$) commute, the equations
(\ref{hamsys}) have a common set of solutions depending on all the evolution
parameters $t_{i}$%
\[
\lambda_{i}=\lambda_{i}(t_{1}=x,t_{2},\ldots,t_{n})\text{, }\mu_{i}=\mu
_{i}(t_{1}=x,t_{2},\ldots,t_{n})\text{.}%
\]
We have, due to (\ref{hamsys}), that
\[
\lambda_{x}\equiv\lambda_{t_{1}}=\frac{\partial H_{1}^{(k,m)}}{\partial\mu
}=G^{(m)}\mu\ \ \text{so that}\ \ \mu=g^{(m)}\lambda_{x},
\]
where the inverse of $G^{(m)}$ (i.e. respective covariant metric tensors) is
denoted as $g^{(m)}$. Observe that the above relation does not depend on $k$.
Using this to eliminate $\mu$ from the first part of (\ref{hamsys}) we obtain
\[
\lambda_{t_{r}}=\frac{\partial H_{r}^{(k,m)}}{\partial\mu}=K_{r}G^{(m)}\mu,
\]
or, according to the above%
\begin{equation}
\lambda_{t_{r}}=K_{r}\lambda_{x}\equiv Z_{r}^{n}(\lambda,\lambda_{x})\text{,
\ \ }r=1,\ldots,n. \label{Fer}%
\end{equation}
This is a set of $n$ autonomous systems of $n$ coupled first order PDE's of
evolutionary type, with the right hand sides depending linearly on the
derivatives $\lambda_{x}$ \cite{blama}. More precisely, it is a set of $n$
integrable dispersionless equations, belonging to the class of so-called
weakly nonlinear semi-Hamiltonian systems \cite{tsar1},\cite{tsar2}, where the
variables $\lambda_{i}$ are the Riemann invariants for (\ref{Fer}). We will
call them \emph{Killing dispersionless system} as they are constructed
directly from Killing tensors.

We will interpret the right hand sides of (\ref{Fer}) as vector fields on an
infinite dimensional manifold $\mathcal{M}$ the points of which are vector
functions of $x$ of the form $u=(\lambda_{1}(x),\ldots,\lambda_{n}(x))$, where
we assume that the functions $\lambda_{i}(x)$ are either periodic in $x$ or
they vanish together with all their derivatives when $x\rightarrow\pm\infty$.
\ A vector field $X$ is at a point $u\in\mathcal{M}$ \ given by an $n$-tuple
of the form $X(u)=(f_{1}[\lambda],\ldots,f_{n}[\lambda])$ where $f_{i}%
[\lambda]=f_{i}(\lambda_{1},\lambda_{1,x},\ldots,\lambda_{2},\lambda
_{2,x},\ldots,\lambda_{n},\lambda_{n,x},\ldots)$ are differential functions of
$\lambda$. Similarly, a covector field $\alpha$ on $\mathcal{M}$ is in a point
$u=(\lambda_{1}(x),\ldots,\lambda_{n}(x))$ given by $\alpha(u)=(g_{1}%
[\lambda],\ldots,g_{n}[\lambda])$. The dual map between $T_{u}\mathcal{M}$ and
$T_{u}^{\ast}\mathcal{M}$ is given by%
\[
\left\langle \alpha,X\right\rangle (u)=\int_{x}%
{\textstyle\sum_{i=1}^{n}}
f_{i}[\lambda]\,g_{i}[\lambda]\,dx.
\]
Here and below the integration is performed over one period (in case of
periodic boundary conditions) or over $\mathbf{R}$ in case of functions
vanishing at $\pm\infty$. All functions and expressions are always assumed to
be integrable. For any two given vector fields $X$ and $Y$ on $\mathcal{M}$
their commutator is defined in a usual way as $\left[  X,Y\right]  =X^{\prime
}[Y]-Y^{\prime}[X]$ where $X^{\prime}[Y]$ denotes the directional derivative
of $X$ in the direction of $Y.$

As was shown in \cite{ferap}, the vector fields $Z_{i}^{n}$ pairwise commute:%
\[
\left[  Z_{i}^{n},Z_{j}^{n}\right]  =0\ \ \ \ \ \ \ \ \ \ \ i,j=1,\ldots,n,
\]
thus, (\ref{Fer}) is a set of $n$ commuting evolutionary dynamic systems
(vector fields) on $\mathcal{M}$. We will need the superscript $n$ to indicate
the number of components (dimension) of these systems. \ Below we will
introduce invariants on (\ref{Fer}) \ that eventually turn these systems into
hierarchies of soliton systems with lower number of fields. This is the main
idea of this paper.

We begin by defining the following differential functions (currents,
'Lagrangians'):%
\begin{equation}
\mathcal{L}_{r}^{(n,m,k)}\overset{\text{def}}{=}\frac{1}{2}\lambda_{x}%
^{T}g^{(m)}K_{r}\lambda_{x}-V_{r}^{(k)}\text{, }r=1,\ldots,n. \label{Lag}%
\end{equation}
In our further considerations we will especially need the first current
$\mathcal{L}_{1}^{(n,m,k)},$ so we will denote it simply by $\mathcal{L}%
^{(n,m,k)},$%
\[
\mathcal{L}^{(n,m,k)}\overset{\text{def}}{=}\mathcal{L}_{1}^{(n,m,k)}%
\]
This current is a Legendre transform of $H_{1}^{(n,m,k)}$ (this is not true
for $H_{r}^{(n,m,k)}$ with $r>1$). These differential functions yield the
following functionals on $\mathcal{M}$:%

\[
I_{r}^{(n,m,k)}(u)\overset{\text{def}}{=}\int_{x}\mathcal{L}_{r}%
^{(n,m,k)}[\lambda]\,dx,
\]
where, as usual, $u=(\lambda_{1}(x),\ldots,\lambda_{n}(x))$. We have, of
course,
\[
\frac{dI_{r}^{(n,m,k)}}{dt_{s}}=\left\langle \frac{\delta I_{r}^{(n,m,k)}%
}{\delta\lambda},Z_{s}^{n}[\lambda]\right\rangle =\left\langle E\left(
\mathcal{L}_{r}^{(n,m,k)}\right)  ,Z_{s}^{n}[\lambda]\right\rangle
\]
where $E=\left(  E_{1},\ldots,E_{n}\right)  =\left(  \frac{\delta}%
{\delta\lambda_{1}},\ldots,\frac{\delta}{\delta\lambda_{n}}\right)  $ is the
Euler-Lagrange operator on $\mathcal{M}$.

\begin{lemma}
\label{niezm}In the notation as above,%
\[
\frac{dI_{1}^{(n,m,k)}}{dt_{r}}=0\text{, \ }r=1,\ldots,n.
\]

\end{lemma}

\begin{proof}
It suffices to prove that%
\begin{equation}
\sum_{i=1}^{n}E_{i}\left(  \mathcal{L}^{(n,m,k)}\right)  \lambda_{i,t_{r}%
}=\sum_{i=1}^{n}E_{i}\left(  \mathcal{L}_{r}^{(n,m,k)}\right)  \lambda_{i,x}
\label{wzorek}%
\end{equation}
since integrating of (\ref{wzorek}) yields,
\[
\frac{dI_{1}^{(n,m,k)}}{dt_{r}}=\frac{dI_{r}^{(n,m,k)}}{dt_{1}}\text{,
\ }r=1,\ldots,n
\]
while
\[
\frac{dI_{r}^{(n,m,k)}}{dt_{1}}=\int_{x}%
{\textstyle\sum_{i=1}^{n}}
E_{i}\left(  \mathcal{L}_{r}^{(n,m,k)}\right)  \lambda_{i,x}\,dx=\int_{x}%
\frac{d}{dx}\left(  \mathcal{L}_{r}^{(n,m,k)}\right)  \,dx=0.
\]
due to the appropriately chosen boundary conditions. The proof of
(\ref{wzorek}) can be found in Appendix A.
\end{proof}

\begin{corollary}
Lemma \ref{niezm}, due to theorem of \cite{lax} (see also \cite{mokhov})
implies that the $2n$-dimensional set $\mathcal{E}\subset\mathcal{M}$
\ defined as%
\[
\mathcal{E}=\left\{  u:E_{i}\left(  \mathcal{L}^{(n,m,k)}\right)  =0\text{ for
all }i=1,\ldots,n\right\}
\]
is $Z_{r}^{n}$-invariant for all $r=1,\ldots,n$.
\end{corollary}

Thus, if $u_{0}\in\mathcal{E}$ then the integral (Fr\"{o}benius)
$n$-dimensional submanifold $\mathcal{S}_{u_{0}}$ of $\mathcal{M}$ spanned by
the commuting vector fields $Z_{r}^{n}$ and containing $u_{0}$ is a subset of
$\mathcal{E}$. This means that the solution $\lambda(x,t_{r})$ of the $r$-th
Killing system in (\ref{Fer}) that starts at a point $u_{0}\in\mathcal{E}$,
i.e. initially satisfying the set of Euler-Lagrange equations
\begin{equation}
E_{i}\left(  \mathcal{L}^{(n,m,k)}\right)  =0\text{, \ }i=1,\ldots,n
\label{EL}%
\end{equation}
remains in $\mathcal{E}$. i.e. always satisfy (\ref{EL}). This further means
that we can use the set of equations (\ref{EL}) to eliminate some of the
variables $\lambda_{i}$ in (\ref{Fer}). Such an operation does not alter
(\ref{Fer}), but reparametrizes it, leading to fewer equations of higher
order, and the dispersion will occur. As we will see below, this operation of
elimination of variables from (\ref{Fer}) through the use of (\ref{EL}) will
lead both to known and new soliton hierarchies in $(1+1)$ dimensions.

\section{Benenti class of St\"{a}ckel systems}

In the rest of this paper we consider the simplest class of separation curves
(\ref{SC1}) in the form%
\begin{equation}
H_{1}\lambda^{n-1}+H_{2}\lambda^{n-2}+\cdots+H_{n}=\frac{1}{2}\lambda^{m}%
\mu^{2}+\lambda^{k} \label{SC}%
\end{equation}
($\lambda,\mu\in\mathbf{R}$ for a moment), where $n\in\mathbf{N}$ while
$m,k\in\mathbf{Z}$. This object contains a complete information about the
so-called Benenti systems \cite{be}-\cite{be1}. Hamiltonian functions
calculated from the related system of separation relations take the form
(\ref{Ham}) \cite{mac2005}. Due to a special form of (\ref{SC}) it turns out
that the metric tensors $G^{(m)}$ are now%
\[
G^{(m)}=L^{m}G^{(0)}\text{, \ with }G^{(0)}=\text{diag}\left(  \frac{1}%
{\Delta_{1}},\ldots,\frac{1}{\Delta_{n}}\right)  \text{, \ }m\in
\mathbf{Z}\text{,}%
\]
where $\Delta_{i}=%
{\textstyle\prod\limits_{j\neq i}}
(\lambda_{i}-\lambda_{j})$ and where $L=$diag$(\lambda_{1},\ldots,\lambda
_{n})$ is a $(1,1)$-tensor on $\mathcal{Q}$ (it is a conformal Killing tensor
with respect to all the metrics $G^{(m)}$). Moreover, Killing tensors $K_{r}$
can now be obtained by the following recursion relation:%
\begin{equation}
K_{r+1}=LK_{r}+q_{r}I\text{, \ }K_{1}=I\text{, \ }K_{n+1}=0\text{, }%
r=1,\ldots,n, \label{Krec}%
\end{equation}
so that indeed they are diagonal (in $\lambda$-coordinates) in accordance with
(\ref{Kdiag}): $K_{r}=$diag$(v_{r}^{1},\ldots,v_{r}^{n})$. The functions
$q_{r}=q_{r}(\lambda)$ are coefficients of the characteristic polynomial of
the tensor $L$ i.e. they are defined by%
\begin{equation}
\det(\lambda I-L)=\sum\limits_{i=0}^{n}q_{i}\lambda^{n-i}, \label{Viete}%
\end{equation}
so that $q_{0}=1,q_{1}=-\sum_{i=1}^{n}\lambda_{i},\ldots,q_{n}=(-1)^{n}%
{\textstyle\prod_{i=1}^{n}}
\lambda_{i}$ ($q_{i}$ are Vi\`{e}te polynomials in the variables $\lambda$).
Moreover, the potentials $V_{r}^{(k)}$ in Hamiltonians (\ref{Ham}) can now be
obtained from the following recursion relation \cite{mac2005}:%
\begin{equation}
V_{r}^{(k)}=V_{r+1}^{(k-1)}-q_{r}V_{1}^{(k-1)}\text{, }\ k\in\mathbf{Z}
\label{rekup}%
\end{equation}
(with the convention that $V_{r}^{(k)}\equiv0$ for$\ \ r<1\ \ $or$\ \ r>n\ $)
with the initial condition:%
\begin{equation}
V_{r}^{(0)}=\delta_{r,n}\text{, \ \ \ \ }r=1,\ldots,n. \label{in}%
\end{equation}
This recursion can be reversed. The inverse recursion is given by%

\begin{equation}
V_{r}^{(k)}=V_{r-1}^{(k+1)}-\frac{q_{r-1}}{q_{n}}V_{n}^{(k+1)}\text{,}%
\ \ k\in\mathbf{Z}\text{, }r=1,\ldots,n\text{.} \label{rekdown}%
\end{equation}
The first potentials are rather trivial:%

\begin{equation}
V_{r}^{(k)}=\delta_{r,n-k}\ \ \ \text{for }k=0,1,...,n-1,\ \ V_{r}%
^{(n)}=-q_{r},\ \ V_{r}^{(-1)}=-\frac{q_{r-1}}{q_{n}}, \label{first}%
\end{equation}
but for $r<-1$ or for $r>n$ the potentials become complicated polynomial (for
$r>n$) or rational (for $r<-1$) functions of $q$.

From (\ref{rekup}) we get
\begin{equation}
V_{n}^{(k)}=-q_{n}V_{1}^{(k-1)}\text{,}\ \ k\in\mathbf{Z} \label{5}%
\end{equation}
and%
\begin{equation}
V_{r}^{(k)}=-q_{r}V_{1}^{(k-1)}-q_{r+1}V_{1}^{(k-2)}-...-q_{n}V_{1}%
^{(k-n+r-1)},\ \ k\in\mathbf{Z} \label{6}%
\end{equation}
while iteration of (\ref{rekdown}) leads to%
\begin{align}
V_{r}^{(k)}  &  =-\frac{1}{q_{n}}\left(  q_{r-1}V_{n}^{(k+1)}+...+q_{1}%
V_{n}^{(k+r-1)}+V_{n}^{(k+r)}\right) \nonumber\\
&  =q_{r-1}V_{1}^{(k)}+...+q_{1}V_{1}^{(k+r-2)}+V_{1}^{(k+r-1)},\ \ k\in
\mathbf{Z}\text{.} \label{7}%
\end{align}

\section{Killing systems and related invariants for the Benenti class in
Vi\`{e}te coordinates}

The functions $q_{r}(\lambda)$ defined in (\ref{Viete}) can serve as a new set
of variables on $\mathcal{Q}$ (we will call them Vi\`{e}te coordinates). It
turns out that these coordinates (that also reparametrize the
infinite-dimensional manifold $\mathcal{M}$ so that $\mathcal{M}\ni
u=(q_{1}(x),\ldots,q_{n}(x))$ now) are much more convenient for our further
purposes. The above considerations, in particular Lemma \ref{niezm} and the
corollary that follows, remain true independently of coordinate system since
the Euler-Lagrange equations are invariant with respect to point
transformations. In this section we sort out the structure of (\ref{Fer}) and
(\ref{EL}) for the Benenti class in Vi\`{e}te coordinates as well as we prove
many other important relations.

The tensors $L$, $L^{-1}$, $G^{(0)}$and $g^{(0)}=\left(  G^{(0)}\right)
^{-1}$ have in Vi\`{e}te coordinates (\ref{Viete}) the form:
\begin{equation}
L=\left(
\begin{array}
[c]{cccc}%
-q_{1} & 1 &  & 0\\
-q_{2} & 0 & \ddots & \\
\vdots &  &  & 1\\
-q_{n} & 0 & \cdots & 0
\end{array}
\right)  ,\ \ \ \ L^{-1}=\left(
\begin{array}
[c]{cccc}%
0 & \cdots & 0 & -\frac{1}{q_{n}}\\
1 & 0 & 0 & -\frac{q_{1}}{q_{n}}\\
& \ddots & 0 & \vdots\\
0 &  & 1 & -\frac{q_{n-1}}{q_{n}}%
\end{array}
\right)  , \label{L}%
\end{equation}%
\begin{equation}
G^{(0)}=\left(
\begin{array}
[c]{cccc}%
0 & 0 & 0 & 1\\
0 & \cdots & \cdots & q_{1}\\
0 & 1 & \cdots & \vdots\\
1 & q_{1} & \cdots & q_{n-1}%
\end{array}
\right)  ,\ \ \ \ g^{(0)}=\left(
\begin{array}
[c]{cccc}%
V_{1}^{(2n-2)} & \cdots & -q_{1} & 1\\
\cdots & \cdots & \cdots & 0\\
-q_{1} & 1 & \cdots & \\
1 & 0 &  & 0
\end{array}
\right)  . \label{G}%
\end{equation}
so that $L_{j}^{i}=V_{i}^{(n-j+1)}$ and $g_{ij}^{(0)}=V_{1}^{(2n-i-j)}$.
Moreover, for the Benenti class, the system (\ref{Fer}) attains in Vi\`{e}te
coordinates (\ref{Viete}) the form $q_{t_{r}}=K_{r}(q)q_{x}$ or, explicitly
\begin{equation}
\frac{d}{dt_{r}}q_{j}=(q_{j+r-1})_{x}+\sum_{k=1}^{j-1}\left(  q_{k}\left(
q_{j+r-k-1}\right)  _{x}-q_{j+r-k-1}\left(  q_{k}\right)  _{x}\right)
\equiv\left(  Z_{r}^{n}\left[  q\right]  \right)  ^{j}\text{ \ \ \ \ }%
r,j=1,\ldots,n \label{Ferq}%
\end{equation}
where $q_{\alpha}=0$ as soon as $\alpha>n$ and $\left(  Z_{r}^{n}\left[
q\right]  \right)  ^{j}$ denotes the $j$-th component of the vector field
$Z_{r}^{n}\left[  q\right]  $. One proves (\ref{Ferq}) by a direct
calculation, using (\ref{L}) and (\ref{Krec}). Observe, that the following
symmetry relation takes place: $\left(  Z_{i}^{n}\left[  q\right]  \right)
^{j}=\left(  Z_{j}^{n}\left[  q\right]  \right)  ^{i}$, \ \ $i,j=1,\ldots,n$.

We can, in accordance with the above, also define the \emph{infinite Killing
hierarchy }for the Benenti class%
\begin{equation}
\frac{d}{dt_{r}}q_{j}=(q_{j+r-1})_{x}+\sum_{k=1}^{j-1}\left(  q_{k}\left(
q_{j+r-k-1}\right)  _{x}-q_{j+r-k-1}\left(  q_{k}\right)  _{x}\right)
\equiv\left(  Z_{r}^{\infty}\left[  q\right]  \right)  ^{j}\text{
\ \ \ }r,j=1,\ldots,\infty\label{Ferinf}%
\end{equation}
that is formally given by the same expression as (\ref{Ferq}) but where we now
do not impose the restriction $q_{\alpha}=0$ for $\alpha>n$. By comparing
(\ref{Ferq}) and (\ref{Ferinf}) one sees directly that for the $r$-th Killing
vector field $Z_{r}^{n}\left[  q\right]  $ from (\ref{Ferq}) its first $n+1-r$
components coincide with the corresponding components of the infinite vector
field $Z_{r}^{\infty}\left[  q\right]  $:%

\begin{equation}
\left(  Z_{r}^{n}\left[  q\right]  \right)  ^{j}=\left(  Z_{r}^{\infty}\left[
q\right]  \right)  ^{j}\text{ \ for }j+r-1\leq n. \label{red}%
\end{equation}

\begin{lemma}
\bigskip The infinite-component vector fields $Z_{r}^{\infty}\left[  q\right]
$ in (\ref{Ferinf}) mutually commute:%
\[
\left[  Z_{i}^{\infty}\left[  q\right]  ,Z_{j}^{\infty}\left[  q\right]
\right]  =0\text{ for all }i,j=1,\ldots\infty
\]

\end{lemma}

\begin{proof}
This can be proved by using $\left[  Z_{i}^{n}\left[  q\right]  ,Z_{j}%
^{n}\left[  q\right]  \right]  =0$ for all $i,j=1,\ldots n$ and (\ref{red}).
Indeed, from (\ref{red}) and the relation
\[
\left(  Z_{i}^{\infty}[q]\right)  ^{j}=\left(  Z_{i}^{\infty}\right)
^{j}[q_{1},...,q_{i+j-1}]
\]
one finds that
\[
\left(  \left[  Z_{i}^{\infty}[q],Z_{j}^{\infty}[q]\right]  \right)
^{l}=\left(  \left[  Z_{i}^{3(n-1)}[q],Z_{j}^{3(n-1)}[q]\right]  \right)
^{l},\ \ \ \ \ \ \ i,j,l=1,...,n
\]
for arbitrary $n\in\mathbf{N}$.
\end{proof}

Let us point out that the infinite Killing hierarchy (\ref{Ferinf}) is exactly
the so-called universal hierarchy considered recently in \cite{al1},\cite{al2}
from the point of view of Lax representation.

\begin{lemma}
\label{1}In Vi\`{e}te coordinates the following relations hold:
\end{lemma}

\begin{enumerate}
\item
\begin{equation}
\frac{\partial V_{1}^{(k)}}{\partial q_{i}}=\frac{\partial V_{1}^{(k+\alpha)}%
}{\partial q_{i+\alpha}}\text{ for }i=1,\ldots,n-\alpha\text{, }k\in
\mathbf{Z}\text{ } \label{8a}%
\end{equation}

\item
\begin{equation}
\left(  L^{k}\right)  _{j}^{i}=V_{i}^{(n+k-j)},\ \ \ \ \ k\in\mathbf{Z}
\label{Lkij}%
\end{equation}

\item
\begin{equation}
g_{ij}^{(m)}=V_{1}^{(2n-m-i-j)},\ \ \ \ \ m\in\mathbf{Z}. \label{gmij}%
\end{equation}

\end{enumerate}

\begin{proof}
For relation (\ref{8a}) the proof is inductive with the help of formula
(\ref{6}). For relation (\ref{Lkij}) the proof is by induction with respect to
$k$. By (\ref{L}), \ $L_{j}^{i}=V_{i}^{(n-j+1)}$. By the induction assumption
and due to the recursion (\ref{rekup})
\[
\left(  L^{k+1}\right)  _{j}^{i}=\sum_{r=1}^{n}L_{r}^{i}\left(  L^{k}\right)
_{j}^{r}=-q_{i}V_{1}^{(n+k-j)}+V_{i+1}^{(n+k-j)}=V_{i}^{(n+k-j+1)}%
\]
which concludes the inductive step up. Similarly, due to the recursion
(\ref{rekdown}),$\ \ $
\[
\left(  L^{k-1}\right)  _{j}^{i}=\sum_{r=1}^{n}\left(  L^{-1}\right)  _{r}%
^{i}\left(  L^{k}\right)  _{j}^{r}=V_{i-1}^{(n+k-j)}-\frac{q_{i-1}}{q_{n}%
}V_{n}^{(n+k-j)}=V_{i}^{(n+k-j-1)}\text{.}%
\]
which concludes the inductive step down. Finally, for relation (\ref{gmij}),
according to (\ref{G}), we have $g_{ij}^{(0)}=V_{1}^{(2n-i-j)}.$ By induction
\[
g_{ij}^{(m+1)}=\sum_{k=1}^{n}V_{1}^{(2n-m-i-k)}\left(  L^{-1}\right)  _{j}%
^{k}.
\]
Thus, due to (\ref{L}) we have for $j<n$
\[
g_{ij}^{(m+1)}=g_{i,j+1}^{(m)}=V_{1}^{(2n-m-i-j-1)}.
\]
while for $j=n$ we have
\begin{align*}
g_{in}^{(m+1)}  &  =-\frac{1}{q_{n}}\left(  q_{n-1}V_{1}^{(n-m-i)}%
+...+q_{1}V_{1}^{(2n-m-2-i)}+V_{1}^{(2n-m-1-i)}\right) \\
&  =-\frac{1}{q_{n}}V_{n}^{(n-m-i)}=V_{1}^{(n-m-i-1)},
\end{align*}
which follows from (\ref{5}) and (\ref{7}). This concludes the inductive step
up. The induction down (for $m<0$) is proved in a similar way.
\end{proof}

The next theorem describes symmetry properties of functions (\ref{Lag}).
Observe that due to (\ref{first}) the functions (\ref{Lag}) are in the Benenti
case geodesic (without the potential part) for $k=0,\ldots,n-1$.

\begin{theorem}
\label{lemma5}For the Lagrangian densities
\[
\mathcal{L}^{n,m,k}=\frac{1}{2}\sum_{i,j=1}^{n}q_{i,x}g_{ij}^{(m)}%
(q)q_{j,x}-V_{1}^{(k)}=\frac{1}{2}\sum_{i,j=1}^{n}q_{i,x}V_{1}^{(2n-m-i-j)}%
q_{j,x}-V_{1}^{(k)}%
\]
the following relations hold:
\end{theorem}

\begin{enumerate}
\item for $\alpha=1,\ldots,n-1$%
\begin{equation}
E_{i}\left(  \mathcal{L}^{n,m,k}\right)  =E_{i-\alpha}\left(  \mathcal{L}%
^{n,m+\alpha,k-\alpha}\right)  ,\ \ \ \ \ \ \ i=\alpha+1,...,n, \label{wtyl}%
\end{equation}

that can also be written as%
\begin{equation}
E_{i}\left(  \mathcal{L}^{n,m,k}\right)  =E_{i+\alpha}\left(  \mathcal{L}%
^{n,m-\alpha,k+\alpha}\right)  ,\ \ \ \ \ \ \ i=1,...,n-\alpha. \label{wprzod}%
\end{equation}

\item
\begin{equation}
E_{l}\left(  \mathcal{L}^{n,0,2n+\sigma}\right)  =E_{l+1}\left(
\mathcal{L}^{n+1,0,2n+\sigma+2}\right)  ,\ \ \ \sigma=1,...,n-1,\ \ \ l=\sigma
+1,...,n. \label{19a}%
\end{equation}

\item
\begin{equation}
E_{l}(\mathcal{L}^{n,n-\sigma,0})=E_{l}\left(  \mathcal{L}^{n+1,n+1-\sigma
,0}\right)  ,\ \ \ \sigma=1,...,n-1,\ \ \ l=1,...,\sigma\label{19a1}%
\end{equation}%
\begin{equation}
E_{\sigma+l}(\mathcal{L}^{n,n-\sigma,-n})_{|q_{j}\rightarrow q_{j+1}}%
=E_{l}\left(  \mathcal{L}^{n+1,n+1-\sigma,-n-1}\right)  ,\ \ l=1,...,n-\sigma
,\ j=1,...,n \label{19a2}%
\end{equation}

\end{enumerate}

The proof of this theorem can be found in Appendix B. As it will be shown in
the next section, Theorem \ref{lemma5} guarantee that the form of invariants
survives the passage from $n$-component to $(n+1)$-component Killing system
and hence it will be crucial for the construction of soliton hierarchies. The
index $\sigma$ will be related with the number of components of the obtained
soliton systems.

\section{Elimination procedure}

\thinspace Using the results of the previous section we will now construct in
a systematic way soliton hierarchies related to Benenti class of separation
relations. These hierarchies will be generated by a procedure of elimination
of variables in the set of dispersionless systems (\ref{Ferq}) with the help
of Euler-Lagrange equations (\ref{EL}) (with suitable chosen parameters $n$,
$m$ and $k$ determining the metric tensor $g^{(m)}$ and the separable
potential $V_{1}^{(k)}$). Actually, we present two separate elimination
procedures, one for positive potentials (i.e. with $k>n$) and one for negative
potentials (i.e. those with $k<0$), leading to different soliton hierarchies.

As we pointed out, the set $\mathcal{E\subset M}$ of solutions of
Euler-Lagrange equations (\ref{EL}) is invariant with respect to all the
vector fields $Z_{r}$ of Killing systems (\ref{Fer}). The same must be true
even in Vi\`{e}te coordinates: (\ref{EL}) written in Vi\`{e}te coordinates is
also invariant with respect to (\ref{Ferq}). This means that we can use the
set of equations (\ref{EL}) to eliminate some of the variables $q_{i}$ in
(\ref{Ferq}), since along the solutions of systems from (\ref{Ferq}) they are
all the time mutually related by the relations (\ref{EL}).

\subsection{Elimination for positive potentials}

First, let us concentrate on the case with positive (polynomial) separable
potentials. Our aim is to produce $s$ ($s\in\mathbf{N}$) commuting $\sigma
$-component ($\sigma\in\mathbf{N}$) vector fields (evolutionary systems) from
(\ref{Ferq}) and (\ref{EL}). In order to do this we choose $n$ as
$n=s+\sigma-1$ and consider the set of systems (\ref{Ferq}) with this chosen
$n$:%
\begin{equation}
q_{t_{r}}=Z_{r}^{s+\sigma-1}\left[  q_{1,}\ldots,q_{s+\sigma-1}\right]
\text{, \ }r=1,\ldots,n=s+\sigma-1. \label{Fers+p-1}%
\end{equation}
Notice that $r$ can reach $r=s+\sigma-1$ but only up to $r=s$ the first
$\sigma$ components of (\ref{Fers+p-1}) are complete in the sense of the
infinite Killing hierarchy (\ref{Ferinf}), i.e. coincide with the
corresponding components of this hierarchy (see (\ref{red})). We will now use
(\ref{EL}) generated by $\mathcal{L}^{n,0,2n+\sigma}$ in order to perform the
elimination. The structure of these equations is described in the lemma below.

\begin{lemma}
\label{elim1}The last $n-\sigma$ invariant equations in (\ref{EL}) for
$\mathcal{L}^{n,0,2n+\sigma}$, with $n=s+\sigma-1$ (so that $m=0$ and
$k=2n+\sigma=2s+3\sigma-2$) have the form
\end{lemma}

\begin{equation}%
\begin{array}
[c]{l}%
w_{\sigma+1}^{\left(  n,\sigma\right)  }\equiv-2q_{n}+\varphi_{\sigma
+1}^{(n,\sigma)}[q_{1},...,q_{n-1}]=0\\
\vdots\\
w_{n}^{\left(  n,\sigma\right)  }\equiv-2q_{\sigma+1}+\varphi_{n}^{(n,\sigma
)}[q_{1},...,q_{\sigma}]=0
\end{array}
\text{\ } \label{ELwprost}%
\end{equation}
where we denoted, to shorten the notation, $E_{i}\left(  \mathcal{L}%
^{n,0,2n+\sigma}\right)  $ as $w_{i}^{(n,\sigma)}$.

\begin{proof}
From the recursion (\ref{rekup}) it follows that
\begin{equation}
V_{1}^{(n+j)}=V_{1}^{(n+j)}(q_{1},...,q_{j+1}),\ \ \ \ j=0,...,n-1 \label{16a}%
\end{equation}
so that, again by (\ref{8a}) and (\ref{6})
\begin{equation}
\frac{\partial V_{1}^{(2n+\sigma)}}{\partial q_{n+1-j+\sigma}}=\frac{\partial
V_{1}^{(n+j)}}{\partial q_{1}}=2q_{j}+f_{j}(q_{1},...,q_{j-1}%
)\ \ \ \ \ \ \ \ j=2,...,n,\ \ \sigma=1,...,n-1 \label{16b}%
\end{equation}
(for $j=1$ we would have $f_{1}\equiv0$) where the first equality follows by
inserting $i=1$, $k=n+j$ and $\alpha=n-j+\sigma$ in (\ref{8a}) and the second
one from the fact that according to (\ref{first}) and (\ref{6}) $V_{1}%
^{(n+j)}=-q_{j+1}+2q_{1}q_{j}+\varphi_{j}(q_{1},...,q_{j-1})$ for
$j=2,\ldots,n$. On the other hand, for the geodesic Lagrangian density
\[
\mathcal{L}^{n,0,0}=\frac{1}{2}\sum_{i,j=1}^{n}V_{1}^{(2n-i-j)}q_{i,x}q_{j,x}%
\]
from (\ref{16c}), as $V_{1}^{(k)}\neq0$ for $k\geq n-1$ and $\partial
V_{1}^{(k)}/\partial q_{1}\neq0$ for $k\geq n$, we find that
\[
E_{l}\left(  \mathcal{L}^{n,0,0}\right)  =F_{l}[q_{1},...,q_{n-l+1}]\text{,
\ }l=1,\ldots,n.
\]
Since $\mathcal{L}_{1}^{n,0,2n+\sigma}=\mathcal{L}_{1}^{n,0,0}-V_{1}%
^{(2n+\sigma)}$, we obtain (putting $j=n+1-i+\sigma$ in (\ref{16b}))
\[
E_{i}\left(  \mathcal{L}^{n,0,2n+\sigma}\right)  =-2q_{\sigma+n-i+1}%
+\varphi_{i}^{(n,\sigma)}[q_{1},...,q_{\sigma+n-i}],\ \ \ i=1,...,n,
\]
(where as usual we denote $q_{\alpha}=0$ for $\alpha>n$) where
\[
\varphi_{i}^{(n,\sigma)}[q_{1},...,q_{\sigma+n-i}]=F_{i}[q_{1},...,q_{n-i+1}%
]+f_{\sigma+n-i+1}(q_{1},...,q_{\sigma+n-i}).
\]

\end{proof}

Due to their structure, equations (\ref{ELwprost}) make it possible to
successively express (eliminate) the variables $q_{\sigma+1},\ldots
,q_{s+\sigma-1}\equiv q_{n}$ as differential functions of $q_{1}%
,\ldots,q_{\sigma}$:%
\begin{equation}%
\begin{array}
[c]{c}%
q_{\sigma+1}=f_{\sigma+1}^{n}\left[  q_{1},\ldots,q_{\sigma}\right] \\
q_{\sigma+2}=f_{\sigma+2}^{n}\left[  q_{1},\ldots,q_{\sigma}\right] \\
\vdots\\
q_{n}=f_{n}^{n}\left[  q_{1},\ldots,q_{\sigma}\right]  .
\end{array}
\label{elim}%
\end{equation}

Let us first observe that performing the elimination (\ref{elim}) in the
systems (\ref{Fers+p-1}) must lead to $\sigma$-component systems of the form
$\overline{q}_{t_{r}}=\overline{Z}_{r}^{\sigma}\left[  q_{1},\ldots,q_{\sigma
}\right]  $, while for each system in (\ref{Fers+p-1}) the last $s-1$
components turn into some system of differential consequences of
$w_{1}^{(n,\sigma)},\ldots,w_{\sigma}^{(n,\sigma)}$ (and are zero on
$\mathcal{S}_{u_{0}}$ i.e. they are satisfied along any solutions of
$\overline{q}_{t_{r}}=\overline{Z}_{r}^{\sigma}$). Therefore, after this
elimination we obtain
\begin{equation}
\left\{
\begin{array}
[c]{l}%
\overline{q}_{t_{r}}=\overline{Z}_{r}^{\sigma}\left[  q_{1},\ldots,q_{\sigma
}\right]  \text{, \ }\\
0=\varphi_{r}^{i}\left[  w_{1},\ldots,w_{\sigma}\right]  \text{, \ \ }%
i=\sigma+1,\ldots,n
\end{array}
\right.  \text{ \ \ \ }r=1,\ldots,n \label{cel}%
\end{equation}
with $\overline{q}=\left(  q_{1},\ldots,q_{\sigma}\right)  ^{T}$ and
$\varphi_{r}^{i}\equiv q_{i,t_{r}}-(Z_{r}^{n})^{i}.$

\begin{lemma}
\emph{The first }$s$\emph{\ vector fields }$\overline{Z}_{r}^{\sigma}$\emph{in
(\ref{cel}) commute}$:$%
\[
\left[  \overline{Z}_{i}^{\sigma},\overline{Z}_{j}^{\sigma}\right]  =0\text{,
\ }i,j=1,\ldots,s.
\]

\end{lemma}

\begin{proof}
Obviously, in general, for $i,j=1,\ldots,n,$
\[
\left[  \overline{Z}_{i}^{\sigma},\overline{Z}_{j}^{\sigma}\right]
=V_{ij}\left[  w_{1}^{(n,\sigma)},\ldots,w_{\sigma}^{(n,\sigma)}\right]
\]
for some vector fields $V_{ij}$ that vanish on $\mathcal{E}\subset\mathcal{M}$
only. \ Assume for a moment that for $n=s+\sigma-1$ and for some $i,j\leq s$
we have $V_{ij}\left[  w_{1}^{(n,\sigma)},\ldots,w_{\sigma}^{(n,\sigma
)}\right]  \neq0.$ As the vector fields $\overline{Z}_{i}^{\sigma}%
,\overline{Z}_{j}^{\sigma}$ were obtained by the reduction of the complete (in
the sense of the infinite hierarchy) components of $Z_{i}^{n},Z_{j}^{n},$thus
by increasing $n\rightarrow n+\beta,$ we do not change the form of $V_{ij},$
which now has to be expressed by a higher dimension invariants $w_{j}%
^{(n+\beta,\sigma)}$: $V_{ij}=V_{ij}\left[  w_{1}^{(n+\beta,\sigma)}%
,\ldots,w_{\sigma}^{(n+\beta,\sigma)}\right]  .$ But $w_{j}^{(n+\beta,\sigma
)}=w_{j}^{(n+\beta,\sigma)}[q_{1},...,q_{n+\beta}]$ and lower dimensional
invariants $w_{j}^{(n,\sigma)}$ are nonexpressible by the higher dimensional
invariants $w_{i}^{(n+\beta,\sigma)}$, so we get a contradiction.
\end{proof}

We will now show that this procedure leads in fact to an infinite hierarchy of
commuting flows. In order to do this, we will for a moment introduce a new
index so that the vector fields in (\ref{cel}) will be denoted $\overline
{Z}_{r}^{n,\sigma}$ as being obtained by reducing the $n$-component Killing
systems (\ref{Fers+p-1}).

\begin{lemma}
In the above notation%
\[
\overline{Z}_{r}^{n+1,\sigma}=\overline{Z}_{r}^{n,\sigma}\text{ for
}r=1,\ldots,s
\]

\end{lemma}

\begin{proof}
According to (\ref{19a}) we have
\[
w_{\sigma+i+1}^{(n+1,\sigma)}=w_{\sigma+i}^{(n,\sigma)}\text{ for }%
i=1,\ldots,n-\sigma.
\]
Thus, increasing $s$ to $s+1$ and keeping $\sigma$ unaltered (so that $n$
changes to $n+1$) the $n-\sigma$ equations (\ref{elim}) change to $n-\sigma+1$
equations
\begin{align*}
\text{ }q_{\sigma+i}  &  =f_{\sigma+i}^{n+1}[q_{1},...,q_{\sigma}%
]=f_{\sigma+i}^{n}[q_{1},...,q_{\sigma}],\ \ \ i=1,...,n-\sigma\\
q_{n+1}  &  =f_{n+1}^{n+1}[q_{1},...,q_{\sigma}]
\end{align*}
so that the variables $q_{\sigma+1},\ldots,q_{n}$ are expressed by the same
functions of $q_{1},...,q_{\sigma}$ and a new elimination equation for
$q_{n+1}$ appears. Moreover, $\left(  Z_{r}^{n+1}[q]\right)  ^{j}=\left(
Z_{r}^{n}[q]\right)  ^{j}$ for $j=1,\ldots,\sigma$ and $r=1,\ldots,s$. Thus,
replacing $q_{\sigma+i}$ by $f_{\sigma+i}^{n}[q_{1},...,q_{\sigma}]$ in
$\left(  Z_{r}^{n+1}[q]\right)  ^{j}$ and in $\left(  Z_{r}^{n}[q]\right)
^{j}$ yields for $j=1,\ldots,\sigma$ and $r=1,\ldots,s$ the same expression.
But the first operation leads to the reduced vector field $\overline{Z}%
_{r}^{n+1,\sigma}$ while the second - to $\overline{Z}_{r}^{n,\sigma}$.
\end{proof}

Let us now take $s+1$ instead of $s$ (so that $n\rightarrow n+1$) in
(\ref{Fers+p-1}) and (\ref{elim}) and perform the reduction. According to the
above lemma we obtain the following sequence of $s+1$ reduced systems:%
\[
\overline{q}_{t_{r}}=\overline{Z}_{r}^{n+1,\sigma}=\overline{Z}_{r}^{n,\sigma
}\text{ for }r=1,\ldots,s\text{ and }\overline{q}_{t_{n+1}}=\overline{Z}%
_{n+1}^{n+1,\sigma}%
\]
i.e. we obtain the same sequence of $s$ systems as before plus an additional
system at the end of the sequence. Therefore, we see that this procedure leads
to infinite hierarchies of commuting systems, since we can always increase $n$
as much as we please without altering the already obtained systems generated
in previous steps.

The procedure described above can be generalized by using only some part of
the equations in (\ref{ELwprost}) in order to perform the elimination, since
all of these equations are invariant along the flows of our Killing systems.
Namely, we can skip the last $\alpha$ (with $0\leq\alpha\leq n-\sigma-1=s-2$)
equations in (\ref{ELwprost}) and use only the remaining equations (i.e.
$w_{\sigma+1}^{\left(  n,\sigma\right)  }=0$, $w_{\sigma+2}^{\left(
n,\sigma\right)  }=0,\ldots,$ $w_{n-\alpha}^{\left(  n,\sigma\right)  }=0$) to
eliminate $q_{\sigma+\alpha+1},\ldots,q_{n}$ in the Killing systems with
$n=s+\sigma+\alpha-1$:%
\begin{equation}
q_{t_{r}}=Z_{r}^{s+\sigma+\alpha-1}\left[  q_{1,}\ldots,q_{s+\sigma+\alpha
-1}\right]  \text{, \ }r=1,\ldots,n=s+\sigma+\alpha-1. \label{Killingsigma}%
\end{equation}
Thus, the index $\alpha$ indicates how many of the last equations in
(\ref{ELwprost}) we "forget" about. It turns out that the elimination that
follows leads also to hierarchies of commuting equations. To see that, let us
first observe, that this elimination can formally be obtained by performing
the above described procedure with the help of the Lagrangian density
$\mathcal{L}^{n,-\alpha,2n+\sigma+\alpha}$, since according to Theorem
\ref{lemma5} we have
\begin{equation}
E_{i}\left(  \mathcal{L}^{n,-\alpha,2n+\sigma+\alpha}\right)  =E_{i-\alpha
}\left(  \mathcal{L}^{n,0,2n+\sigma}\right)  \equiv w_{i-\alpha}^{(n,\sigma
)}\text{ \ for }i=\alpha+1,\ldots,n\text{.} \label{tesamew}%
\end{equation}
Denoting $E_{i}\left(  \mathcal{L}^{n,m,2n+k}\right)  $ as $w_{i}^{(n,m,k)}$,
where now $w_{i}^{(n,0,k)}\equiv w_{i}^{(n,k)}$ (in the notation of
(\ref{ELwprost})), the last $n-\sigma-\alpha$ Euler-Lagrange equations
(invariants), associated with $\mathcal{L}_{1}^{n,-\alpha,2n+\sigma+\alpha}$,
have therefore the form
\begin{equation}%
\begin{array}
[c]{l}%
w_{\sigma+\alpha}^{(n,-\alpha,\sigma+\alpha)}=w_{\sigma}^{(n,\sigma
)}=w_{\sigma}^{\left(  n,\sigma\right)  }[q_{1},...,q_{n}]=0\\
w_{\sigma+1+\alpha}^{(n,-\alpha,\sigma+\alpha)}=w_{\sigma+1}^{\left(
n,\sigma\right)  }=-2q_{n}+\varphi_{\sigma+1}^{(n,\sigma)}[q_{1}%
,...,q_{n-1}]=0\\
\vdots\\
w_{n}^{(n,-\alpha,\sigma+\alpha)}=w_{n-\alpha}^{\left(  n,\sigma\right)
}=-2q_{\sigma+\alpha+1}+\varphi_{n-\alpha}^{(n,\sigma)}[q_{1},...,q_{\sigma
+\alpha}]=0.
\end{array}
\label{EL1}%
\end{equation}
These equations make it possible to successively express (eliminate) the
variables $q_{\sigma+\alpha+1},\ldots,q_{n}$ as differential functions of
$q_{1},\ldots,q_{\sigma+\alpha}$, which yields%
\begin{equation}%
\begin{array}
[c]{l}%
q_{\sigma+\alpha+1}=q_{\sigma+\alpha+1}\left[  q_{1},\ldots,q_{\sigma+\alpha
}\right] \\
\vdots\\
q_{n}=q_{n}\left[  q_{1},\ldots,q_{\sigma+\alpha}\right]  .
\end{array}
\label{eli2}%
\end{equation}
Therefore, after this elimination the Killing equations (\ref{Killingsigma})
take the form
\begin{equation}
\left\{
\begin{array}
[c]{l}%
\overline{q}_{t_{r}}=\overline{Z}_{r}^{\sigma+\alpha}\left[  \overline
{q}\right]  \text{, \ }\\
0=\varphi_{r}^{i}\left[  w_{1}^{(n,-\alpha,\sigma+\alpha)},\ldots
,w_{\sigma+\alpha}^{(n,-\alpha,\sigma+\alpha)}\right]  \text{, \ \ }%
i=\sigma+\alpha+1,\ldots,n
\end{array}
\right.  \text{ \ \ \ }r=1,\ldots,n=s+\sigma+\alpha-1 \label{cel1}%
\end{equation}
with $\overline{q}=\left(  q_{1},\ldots,q_{\sigma+\alpha}\right)  ^{T}$ and
$\varphi_{r}^{i}\equiv q_{i,t_{r}}-(Z_{r}^{n})^{i}$ (so that the reduced
systems will have $N=\sigma+\alpha$ components). Similarly as before, in
Killing equations (\ref{Killingsigma}), only up to $r=s$ the first
$\sigma+\alpha$ components are complete in the sense of the infinite Killing
hierarchy (\ref{Ferinf}). As before, it stems from the fact that the first
$s$\ vector fields $\overline{Z}_{r}^{\sigma+\alpha}$commute to zero$:$%
\[
\left[  \overline{Z}_{i}^{\sigma+\alpha},\overline{Z}_{j}^{\sigma+\alpha
}\right]  =0\text{, \ }i,j=1,\ldots,s.
\]
the proof of which is analogous as in the case $\alpha=0$ but now we have to
take $n=s+\sigma+\alpha-1$. As previously, we can repeat the elimination
procedure taking $s+1$ instead of $s$ (so that $n$ increases to $n+1$ and
$k=2n+\sigma+\alpha$ increases to $2(n+1)+\sigma+\alpha=k+2$ while $\sigma$
and $\alpha$ are kept unaltered). By the same argument as before, this new
procedure (with $n+1$ instead of $n$) will lead to a sequence of $s+1$
autonomous $(\sigma+\alpha)$-component systems in which the first $s$ systems
will coincide with the corresponding systems obtained from the original
procedure (with $n$). Thus, again we will obtain infinite hierarchies of
soliton systems.

\subsection{Elimination for negative potentials}

We now present the second possibility of elimination - with the use of
negative (rational) separable potentials. Again, our aim is to produce $s$
($s\in\mathbf{N}$) commuting $\sigma$-component ($\sigma\in\mathbf{N}$) vector
fields (evolutionary systems) from (\ref{Ferq}) and (\ref{EL}). This time
however we have to choose $n=s+2\sigma-1$ and the Lagrangian density
$\mathcal{L}^{n,n-\sigma,-n}$ in order to create an infinite hierarchy of
commuting flows.

\begin{lemma}
\label{elim2}The first $n-\sigma$ invariant equations (\ref{EL}) with
$\mathcal{L}^{n,n-\sigma,-n}$, i.e. with $m=n-\sigma$ and $k=-n$, have the
form
\begin{align}
v_{1}^{(n,\sigma)}  &  \equiv-\frac{1}{q_{n}^{2}}+\,\gamma_{1}^{(n,\sigma
)}[q_{1},\ldots,q_{\sigma}]=0,\label{ELn}\\
v_{i}^{(n,\sigma)}  &  \equiv\frac{2q_{n-i+1}}{q_{n}^{3}}\,+\gamma
_{i}^{(n,\sigma)}[q_{1},\ldots,q_{\sigma-i+1},q_{n-i+2},\ldots,q_{n}%
]=0,\ \ i=2,...,n-\sigma.\nonumber
\end{align}
where we denote, to shorten the notation, $E_{i}\left(  \mathcal{L}%
^{n,n-\sigma,-n}\right)  $ as $v_{i}^{(n,\sigma)}$ and $q_{\alpha}=0$ when
$\alpha<1$.
\end{lemma}

\begin{proof}
From the recursion (\ref{rekdown}) it follows that%
\begin{equation}
V_{1}^{(-j)}=V_{1}^{(-j)}(q_{n-j+1},\ldots,q_{n})\text{, }j=1\ldots,n
\label{16aa}%
\end{equation}
From this and from (\ref{do1}), (\ref{16a}) and (\ref{do2}), we have
\[
E_{i}(\mathcal{L}^{n,n-\sigma,0})=G_{i}[q_{1},...,q_{\sigma-i+1}%
],\ \ \ i=1,...,\sigma,
\]%
\begin{equation}
E_{i}(\mathcal{L}^{n,n-\sigma,0})=G_{i}[q_{n-i+1+\sigma},...,q_{n}%
],\ \ \ i=\sigma+1,...,n-\sigma. \label{ELnegproof}%
\end{equation}
Moreover, by using Lemma \ref{1} we find
\[
\frac{\partial V_{1}^{(-n+1-i)}}{\partial q_{1}}=-\frac{2q_{n-i+1}}{q_{n}^{3}%
}+g_{i}(q_{n-i+2},\ldots,q_{n}),\ \ \ i=2,...,n-\sigma
\]
and%
\[
\frac{\partial V_{1}^{(-n+\sigma)}}{\partial q_{\sigma+1}}=\frac{\partial
V_{1}^{(-n)}}{\partial q_{1}}=\frac{1}{q_{n}^{2}}.
\]
Plugging all this into $E_{i}\left(  \mathcal{L}^{n,n-\sigma,-n}\right)
,\ i=1,...,n-\sigma,$ we obtain (\ref{ELn}) where $\gamma_{i}^{(n,\sigma
)}[q]=G_{i}[q]-g_{i}(q)$.
\end{proof}

Let us now consider the following Killing systems%
\begin{equation}
q_{t_{r}}=Z_{r}^{n}[q]\text{, \ \ \ \ }r=\sigma+1,\ldots,\sigma+s\text{,
\ \ \ with\ }n=s+2\sigma-1 \label{nK}%
\end{equation}
We can use the $n-\sigma$ equations (\ref{ELn}) to successively express
(eliminate) the variables $q_{\sigma+1},\ldots,q_{n}$ as differential
functions of $q_{1},\ldots,q_{\sigma}$. This leads to the elimination
relations of the form%
\begin{equation}
q_{\sigma+i}=f_{\sigma+i}^{n}\left[  q_{1},\ldots,q_{\sigma}\right]  \text{,
\ \ }i=1,\ldots,n-\sigma. \label{elimn}%
\end{equation}
Performing the elimination (\ref{elimn}) in (\ref{nK}) we obtain an autonomous
sequence of $s$ evolution equations
\begin{equation}
\overline{q}_{t_{r}}=\overline{Z}_{r}^{\sigma}\left[  q_{1},\ldots,q_{\sigma
}\right]  \text{ \ , \ }r=\sigma+1,\ldots,\sigma+s \label{redn}%
\end{equation}
such that the vector fields $\overline{Z}_{r}^{\sigma}$ mutually commute to
zero. One proves this by the same arguments as in the positive case, since the
first $\sigma$ components of all the vector fields in (\ref{nK}) are complete
in the sense of infinite hierarchy (\ref{Ferinf})).

Analogously to the positive case, we will now show that this procedure leads
to an infinite hierarchy. As in the positive case, we will for a moment
introduce a new index so that the vector fields in (\ref{redn}) will be
denoted $\overline{Z}_{r}^{n,\sigma}$ as being obtained by reducing the
$n$-component Killing systems (\ref{nK}).

\begin{lemma}
\label{cojest}\bigskip In the notation as above%
\[
\overline{Z}_{r+1}^{n+1,\sigma}=\overline{Z}_{r}^{n,\sigma}\text{ for
}r=\sigma+1,\ldots,\sigma+s
\]

\end{lemma}

\begin{proof}
Let us observe that, according to results (\ref{wprzod}), (\ref{19a1}),
(\ref{19a2}) and (\ref{B9})
\begin{equation}
v_{i}^{(n+1,\sigma)}=\left.  v_{i}^{(n,\sigma)}\right\vert _{q_{j}\rightarrow
q_{j+1},\ j=n-i+1,...,n}\text{ for }i=1,\ldots,n-\sigma. \label{numer}%
\end{equation}
Thus, increasing $s$ to $s+1$ and keeping $\sigma$ unaltered (so that $n$
changes to $n+1$) the $n-\sigma$ equations (\ref{elimn}) change to
$n-\sigma+1$ equations
\begin{equation}
q_{\sigma+1}=f_{\sigma+1}^{n+1}[q_{1},...,q_{\sigma}]\text{ \ , \ }%
q_{\sigma+i+1}=f_{\sigma+i+1}^{n+1}[q_{1},...,q_{\sigma}]=f_{\sigma+i}%
^{n}[q_{1},...,q_{\sigma}],\ \ \ i=1,...,n-\sigma. \label{abc}%
\end{equation}
Observe that the last $n-\sigma$ equations (\ref{elimn}) express now
$q_{\sigma+i+1}$ (instead of $q_{\sigma+i}$) as $f_{\sigma+i}^{n}%
[q_{1},...,q_{\sigma}]$. On the other hand, due to (\ref{Ferq}), by changing
$q_{i}\rightarrow q_{i+1}$ for all $i>\sigma$ in the sequence $q_{t_{r}}%
=Z_{r}^{n+1}[q]$ we transform it so that $\left(  Z_{r}^{n+1}[q]\right)
^{j}\rightarrow\left(  Z_{r+1}^{n+1}[q]\right)  ^{j}$ for $j=1,\ldots,\sigma$
and $r=\sigma+1,\ldots,\sigma+s$. Thus, inserting $f_{\sigma+i}^{n}%
[q_{1},...,q_{\sigma}]$ instead of $q_{\sigma+i+1}$ in $\left(  Z_{r+1}%
^{n+1}[q]\right)  ^{j}$ (for $j=1,\ldots,\sigma$ and $r=\sigma+1,\ldots
,\sigma+s$) yields the same expression as inserting the same function
$f_{\sigma+i}^{n}[q_{1},...,q_{\sigma}]$ instead of $q_{\sigma+i}$ in $\left(
Z_{r}^{n+1}[q]\right)  ^{j}$. But the first operation leads to the reduced
vector field $\overline{Z}_{r+1}^{n+1,\sigma}$ while the second - to
$\overline{Z}_{r}^{n,\sigma}$.
\end{proof}

Let us now take $s+1$ instead of $s$ (so that $n\rightarrow n+1$) in
(\ref{nK}) and (\ref{elimn}) and perform the reduction. According to Lemma
\ref{cojest} we obtain the following sequence of $s+1$ reduced systems:%
\[
\overline{q}_{t_{\sigma+1}}=\overline{Z}_{\sigma+1}^{n+1,\sigma}\text{ \ ,
\ }\overline{q}_{t_{r+1}}=\overline{Z}_{r+1}^{n+1,\sigma}=\overline{Z}%
_{r}^{n,\sigma}\text{ for }r=\sigma+1,\ldots,\sigma+s
\]
i.e. we obtain the same sequence of $s$ systems as before but shifted and an
additional system in the beginning of the sequence. This first system can
therefore be treated as \ a next system in some infinite, commuting hierarchy
of vector fields.

Let us also observe that we could use (\ref{elimn}) to eliminate variables in
Killing systems of the form (\ref{Fers+p-1}) and this would lead to a system
of $s$ commuting evolutionary systems. However, this choice does not lead to
any hierarchy: by increasing $s$ to $s+1$ we obtain a different sequence of systems.

As before, this procedure can be generalized: we can use the first
$n-\sigma-\alpha$ equations ($0\leq\alpha<n-\sigma-1$) in (\ref{ELn}) to
eliminate $q_{\sigma+\alpha+1},\ldots,q_{n}$ from the following sequence of
$s$ Killing systems%
\begin{equation}
q_{t_{r}}=Z_{r}^{n}[q]\text{, \ \ \ \ }r=\sigma+\alpha+1,\ldots,\sigma
+\alpha+s\text{ \ with }n=s+2(\sigma+\alpha)-1. \label{Kna}%
\end{equation}
This elimination leads - similarly as above - to $s$ commuting to zero
$N=\sigma+\alpha$ -component systems $\overline{q}_{t_{r}}=\overline{Z}%
_{r}^{\sigma+\alpha}\left[  q_{1},\ldots,q_{\sigma+\alpha}\right]  $ and by
increasing $s$ by $1$ we always obtain a new system of the hierarchy at the
beginning of the sequence.

Next section contains some examples of the above described elimination procedures.

\section{Examples}

\subsection{Elimination with positive potentials}

Below we will present some examples performed with the help of the
(generalized) procedure described in the previous section. Soliton hierarchies
are now classified by pairs $(\sigma,\alpha),$ $\sigma=1,2,...,\alpha
=0,1,...,$ where $N=\sigma+\alpha$ is a number of components in the systems of
a given hierarchy. Assume we would like to construct first $s$ members of the
hierarchy. We have then to fix $n=s+\sigma+\alpha-1$ and take first $s$
Killing equations in (\ref{Killingsigma}). Then, we have to eliminate
coordinates $q_{\sigma+\alpha+1},...,q_{\sigma+\alpha+s-1}=q_{n}$ using
invariants $w_{\sigma+1}^{\left(  n,\sigma\right)  }=0,...,w_{n-\alpha
}^{\left(  n,\sigma\right)  }=0.$ According to (\ref{tesamew}), these
invariants can be generated, for example, from $\mathcal{L}^{n,0,2n+\sigma}$
by taking the equations $E_{i}\left(  \mathcal{L}^{n,0,2n+\sigma}\right)
=0,\ i=\sigma+1,...,n-\alpha.$ After the elimination procedure, soliton
equations are represented by first $N=\sigma+\alpha$ components of first $s$
reduced Killing equations. Observe, that in this procedure the first soliton
equation has always the trivial form $\overline{q}_{t_{1}}=\overline{q}_{x},$
$\overline{q}=\left(  q_{1},\ldots,q_{\sigma+\alpha}\right)  ^{T}.$

Let us start with a one-field hierarchy: $N=\sigma+\alpha=1.$ There is only
one possibility here: $\sigma=1,\alpha=0.$ We present how to produce first
$s=3$ flows which will be recognized as the first members of the KdV
hierarchy. We have therefore to take $n=3$ and $k=7$. Killing systems
(\ref{Ferq}) have the form:%
\begin{align}
\frac{d}{dt_{1}}\left(
\begin{array}
[c]{c}%
q_{1}\\
q_{2}\\
q_{3}%
\end{array}
\right)   &  =\left(
\begin{array}
[c]{c}%
q_{1,x}\\
q_{2,x}\\
q_{3,x}%
\end{array}
\right)  =Z_{1}^{3}\nonumber\\
\frac{d}{dt_{2}}\left(
\begin{array}
[c]{c}%
q_{1}\\
q_{2}\\
q_{3}%
\end{array}
\right)   &  =\left(
\begin{array}
[c]{c}%
q_{2,x}\\
q_{3,x}+q_{1}q_{2,x}-q_{2}q_{1,x}\\
q_{1}q_{3,x}-q_{3}q_{1,x}%
\end{array}
\right)  =Z_{2}^{3}\label{E1}\\
\frac{d}{dt_{3}}\left(
\begin{array}
[c]{c}%
q_{1}\\
q_{2}\\
q_{3}%
\end{array}
\right)   &  =\left(
\begin{array}
[c]{c}%
q_{3,x}\\
q_{1}q_{3,x}-q_{3}q_{1,x}\\
q_{2}q_{3,x}-q_{3}q_{2,x}%
\end{array}
\right)  =Z_{3}^{3}\nonumber
\end{align}
while the Lagrangian $\mathcal{L}^{3,0,7\,}$ is%
\[
\mathcal{L}^{3,0,7\,}=\frac{1}{2}q_{1}^{2}q_{1,x}^{2}-\frac{1}{2}q_{2}%
q_{1,x}^{2}-q_{1}q_{1,x}q_{2x}+q_{1,x}q_{3,x}+\frac{1}{2}q_{2,x}^{2}%
-2q_{2}q_{3}+3q_{1}^{2}q_{3}+3q_{1}q_{2}^{2}-4q_{1}^{3}q_{2}+q_{1}^{5}%
\]
and the Euler-Lagrange equations (\ref{ELwprost}) for the above Lagrangian
attain the form%
\[%
\begin{array}
[c]{l}%
w_{2}^{(3,1)}\equiv-2q_{3}+6q_{1}q_{2}-4q_{1}^{3}+\frac{1}{2}q_{1,x}^{2}%
+q_{1}q_{1,xx}-q_{2,xx}=0\\
w_{3}^{(3,1)}\equiv-2q_{2}+3q_{1}^{2}-q_{1,xx}=0.
\end{array}
\]
These equations can be solved with respect to $q_{2},q_{3}$ yielding
(\ref{elim}) of the form%
\begin{equation}
q_{2}=-\tfrac{1}{2}q_{1,xx}+\tfrac{3}{2}q_{1}^{2}\text{ \ \ , \ \ }%
q_{3}=\tfrac{1}{4}q_{1,xxxx}-\tfrac{5}{2}q_{1}q_{1,xx}-\tfrac{5}{4}q_{1,x}%
^{2}+\tfrac{5}{2}q_{1}^{3}. \label{E2}%
\end{equation}
Substituting it to the above Killing systems gives (\ref{cel}) that read now
explicitly as%
\begin{align*}
&  \left\{
\begin{array}
[c]{l}%
q_{1,t_{1}}=q_{1,x}=\overline{Z}_{1}^{1}\\
0=0\\
0=0
\end{array}
\right.  \text{ \ \ \ \ \ \ \ }\left\{
\begin{array}
[c]{l}%
q_{1,t_{2}}=-\frac{1}{2}q_{1,xxx}+3q_{1}q_{1,x}=\overline{Z}_{2}^{1}\\
0=0\\
0=\frac{1}{2}w_{1,x}%
\end{array}
\right. \\
&  \left\{
\begin{array}
[c]{l}%
q_{1,t_{3}}=\frac{1}{4}q_{1,xxxxx}-5q_{1,x}q_{1,xx}-\frac{5}{2}q_{1}%
q_{1,xxx}+\frac{15}{2}q_{1}^{2}q_{1,x}=\overline{Z}_{3}^{1}\\
0=\frac{1}{2}w_{1,x}\\
0=-\frac{1}{4}w_{1,xxx}+\frac{3}{2}q_{1}w_{1,x}%
\end{array}
\right.
\end{align*}
so that the first components $q_{1,t_{i}}=\overline{Z}_{i}^{1}[q_{1}]$ are the
first three flows of the KdV hierarchy while the remaining equations are just
differential consequences of $w_{1},$ which of course vanish on any
$\mathcal{S}_{u_{0}}$. By taking larger $s$ we can produce an arbitrary number
of flows from the KdV hierarchy.

Next, let us consider two-field systems: $N=\sigma+\alpha=2$. There are two
possibilities: $(\sigma,\alpha)=(2,0)$ and $(\sigma,\alpha)=(1,1).$ Therefore,
as a second example we consider the case $(\sigma,\alpha)=(2,0),$ and $s=3.$
We have now to take $n=s+\sigma+\alpha-1=4$ and $k=2n+\sigma=10$. \ The
Euler-Lagrange equations $E_{4}(\mathcal{L}^{4,0,10\,})=w_{4}^{(4,2)}=0$ and
$E_{3}(\mathcal{L}^{4,0,10\,})=w_{3}^{(4,2)}=0$ can be solved with respect to
$q_{3},q_{4}$ yielding (\ref{elim}) of the form%
\begin{align*}
q_{3}  &  =-\tfrac{1}{2}q_{1,xx}+3q_{1}q_{2}-2q_{1}^{3}\\
q_{4}  &  =\tfrac{1}{4}q_{1,x}^{2}-\tfrac{1}{2}q_{2,xx}-q_{1}q_{1,xx}%
-\tfrac{7}{2}q_{1}^{4}+3q_{1}^{2}q_{2}+\tfrac{3}{2}q_{2}^{2}.
\end{align*}
Substituting it to two first components of Killing equations $Z_{2}%
^{4}[q],\ Z_{3}^{4}[q]$ yields two first nontrivial members of another
two-field soliton hierarchy:%
\begin{equation}%
\begin{array}
[c]{l}%
q_{1,t_{2}}=q_{2,x}\\
q_{2,t_{2}}=-\frac{1}{2}q_{1,xxx}+4q_{1}q_{2,x}+2q_{2}q_{1,x}-6q_{1}%
^{2}q_{1,x}%
\end{array}
, \label{E7}%
\end{equation}
and%
\[%
\begin{array}
[c]{l}%
q_{1,t_{3}}=-\frac{1}{2}q_{1,xxx}+3q_{1}q_{2,x}+3q_{2}q_{1,x}-6q_{1}%
^{2}q_{1,x}\\
q_{2,t_{3}}=-\frac{1}{2}q_{2,xxx}-\frac{3}{2}q_{1}q_{1,xxx}+3q_{2}%
q_{2,x}+6q_{1}q_{2}q_{1,x}+6q_{1}^{2}q_{2,x}-18q_{1}^{3}q_{1,x}.
\end{array}
\]
In the second two-field case $(\sigma,\alpha)=(1,1)$, if we keep $s=3$
unchanged, we have to take $n=s+\sigma+\alpha-1=4$ and $k=2n+\sigma=9$. \ From
the Euler-Lagrange equations (\ref{ELwprost}) for $\mathcal{L}^{4,0,9\,}$
\[
E_{3}(\mathcal{L}^{4,0,9\,})=w_{3}^{(4,1)}=0\text{, \ }E_{2}(\mathcal{L}%
^{4,0,9\,})=w_{2}^{(4,1)}=0
\]
we can eliminate $q_{3}$ and $q_{4}$, which yields (\ref{eli2}). Explicitly,
we obtain:
\begin{align*}
q_{3}  &  =\tfrac{1}{4}q_{1,xxxx}-\tfrac{5}{2}q_{1}q_{1,xx}-\tfrac{5}%
{4}q_{1,x}^{2}+\tfrac{5}{2}q_{1}^{3}\\
q_{4}  &  =\tfrac{1}{4}q_{2,xxxx}-\tfrac{1}{4}q_{1}q_{1,xxxx}-\tfrac{3}%
{4}q_{1,x}q_{1,xxx}-\tfrac{1}{2}q_{1,xx}^{2}-q_{2}q_{1,xx}-\tfrac{5}{2}%
q_{1,x}q_{2,x}\\
&  +4q_{1}^{2}q_{1,xx}-\tfrac{5}{2}q_{1}q_{2,xx}+\tfrac{25}{4}q_{1}q_{1,x}%
^{2}+3q_{1}^{2}q_{2}-\tfrac{7}{2}q_{1}^{4}+\tfrac{3}{2}q_{2}^{2}.
\end{align*}
Then, two first components of the Killing equations $q_{t_{2}}=Z_{2}%
^{4}[q],\ q_{t_{3}}=Z_{3}^{4}[q]$ turn into
\[%
\begin{array}
[c]{l}%
q_{1,t_{2}}=q_{2,x}\\
q_{2,t_{2}}=-\frac{1}{2}q_{2,xxx}+\frac{1}{2}q_{1}q_{1,xxx}+q_{1,x}%
q_{1,xx}+4q_{1}q_{2,x}+2q_{2}q_{1,x}-6q_{1}^{2}q_{1,x}%
\end{array}
,
\]
and%
\[%
\begin{array}
[c]{l}%
q_{1,t_{3}}=-\frac{1}{2}q_{2,xxx}+\frac{1}{2}q_{1}q_{1,xxx}+q_{1,x}%
q_{1,xx}+3q_{1}q_{2,x}+3q_{2}q_{1,x}-6q_{1}^{2}q_{1,x}\\
q_{2,t_{3}}=\frac{1}{4}q_{2,xxxxx}-\frac{1}{4}q_{1}q_{1,xxxxx}-q_{1,x}%
q_{1,xxxx}-\frac{7}{4}q_{1,xx}q_{1,xxx}-q_{2}q_{1,xxx}-\frac{7}{2}%
q_{2,x}q_{1,xx}\\
~~\ \ \ \ \ \ \ \ +\frac{9}{2}q_{1}^{2}q_{1,xxx}-3q_{1}q_{2,xxx}-\frac{9}%
{2}q_{1,x}q_{2,xx}+21q_{1}q_{1,x}q_{1,xx}+6q_{1}q_{2}q_{1,x}+3q_{2}q_{2,x}\\
\ \ \ \ \ \ \ \ \ \ +6q_{1}^{2}q_{2,x}+6q_{1,x}^{2}-18q_{1}^{3}q_{1,x}.
\end{array}
\]
Finally, we shortly mention the three-field case: $N=\sigma+\alpha=3.$ There
are three different hierarchies with the following first nontrivial member of
each hierarchy: \newline for $(\sigma,\alpha)=(3,0):$
\begin{align*}
q_{1,t_{2}}  &  =q_{2,x}\\
q_{2,t_{2}}  &  =q_{3,x}+q_{1}q_{2,x}-q_{2}q_{1,x}\\
q_{3,t_{2}}  &  =-\tfrac{1}{2}q_{2,xxx}-12q_{1}q_{2}q_{1,x}-6q_{1}^{2}%
q_{2,x}+3q_{2}q_{2,x}+2q_{3}q_{1,x}+4q_{1}q_{3,x}+10q_{1}^{3}q_{1,x},
\end{align*}
for $(\sigma,\alpha)=(2,1):$%
\begin{align*}
q_{1,t_{2}}  &  =q_{2,x}\\
q_{2,t_{2}}  &  =q_{3,x}+q_{1}q_{2,x}-q_{2}q_{1,x}\\
q_{3,t_{2}}  &  =-\tfrac{1}{2}q_{2,xxx}+\tfrac{1}{2}q_{1}q_{1,xxx}%
+q_{1,x}q_{1,xx}-12q_{1}q_{2}q_{1,x}-6q_{1}^{2}q_{2,x}+3q_{2}q_{2,x}\\
&  +2q_{3}q_{1,x}+4q_{1}q_{3,x}+10q_{1}^{3}q_{1,x},
\end{align*}
and for $(\sigma+\alpha)=(1,2):$
\begin{align*}
q_{1,t_{2}}  &  =q_{2,x}\\
q_{2,t_{2}}  &  =q_{3,x}+q_{1}q_{2,x}-q_{2}q_{1,x}\\
q_{3,t_{2}}  &  =-\tfrac{1}{2}q_{3,xxx}+\tfrac{1}{2}q_{1}q_{2,xxx}+\tfrac
{1}{2}q_{2}q_{1,xxx}-\tfrac{1}{2}q_{1}^{2}q_{1,xxx}-2q_{1}q_{1,x}%
q_{1,xx}+q_{1,x}q_{2,xx}+q_{2,x}q_{1,xx}\\
&  -\tfrac{1}{2}q_{1,x}^{3}+2q_{3}q_{1,x}+4q_{1}q_{3,x}-12q_{1}q_{2}%
q_{1,x}-6q_{1}^{2}q_{2,x}+3q_{2}q_{2,x}+10q_{1}^{3}q_{1,x}%
\end{align*}
In general, for a fixed $N=\sigma+\alpha,$ this procedure leads to $N$
different $N-$component hierarchies of soliton systems. As the field
representation of constructed hierarchies is non-standard, it is not easy to
recognize which hierarchies are known and which are new. We immediately
recognized the KdV hierarchy. We also found that two-field hierarchy\ starting
from (\ref{E7}) turns after the transformation
\[
u_{1}=-3q_{1}^{2}+2q_{2},\ \ \ u_{2}=2q_{1},\text{ \ }x\rightarrow\sqrt
{2}ix,\text{ \ }t\rightarrow\sqrt{2}it
\]
into the 2-component coupled KdV hierarchy in the representation of Fordy and
Antonowicz \cite{forancoupled}. For example, the first flow of this hierarchy
(\ref{E7}) turns into%
\begin{align*}
u_{1,t_{1}}  &  =\tfrac{1}{4}u_{2,xxx}+\tfrac{1}{2}u_{2}u_{1,x}+u_{1}u_{2,x}\\
u_{2,t_{1}}  &  =u_{1,x}+\tfrac{3}{2}u_{2}u_{2,x}.
\end{align*}
(yielding (3.18) in \cite{forancoupled}).

\subsection{Elimination with negative potentials}

We start by presenting the first two ($s=2$) flows of the only $N=1$-component
hierarchy that can be obtained within our scheme by using the negative
separable potentials. Since $N=1=\alpha+\sigma$, the only choice is to put
$\sigma=1$, $\alpha=0$, which yields $n=s+2(\sigma+\alpha)-1=3$. The
Euler-Lagrange equations (\ref{ELn}) for
\[
\mathcal{L}^{n,n-\sigma,-n}=\mathcal{L}^{3,2,-3}=\frac{1}{2}q_{1,x}^{2}%
-\frac{q_{2,x}q_{3,x}}{q_{3}}+\frac{q_{3,x}^{2}q_{2}}{2q_{3}}-\frac{q_{1}%
}{q_{3}^{2}}+\frac{q_{2}^{2}}{q_{3}^{3}}%
\]
attain the form:%
\begin{equation}
-1-q_{3}^{2}q_{1,xx}=0\text{, \ \ }4q_{2}+2q_{3}^{2}q_{3,xx}-q_{3}q_{3,x}%
^{2}=0 \label{el31-2}%
\end{equation}
which allows for expressing $q_{2}$ and $q_{3}$ as differential functions of
$q_{1}$:%
\begin{align*}
q_{3}  &  =q_{3}[q_{1}]=\left(  -q_{1,xx}\right)  ^{-1/2}\\
q_{2}  &  =q_{2}[q_{1}]=-\tfrac{1}{16}\left(  5q_{1,xxx}^{2}-4q_{1,xx}%
q_{1,xxxx}\right)  \left(  -q_{1,xx}\right)  ^{-7/2}%
\end{align*}
(here and in what follows we only consider the positive solution for $q_{n}$,
otherwise we can change $t\rightarrow-t$). Substituting these expressions to
the Killing systems (\ref{Kna}) and performing the necessary derivations we
obtain the following two commuting flows:%
\[
q_{1,t_{2}}=\left(  q_{2}[q_{1}]\right)  _{x}\text{ \ , \ \ }q_{1,t_{3}%
}=\left(  q_{3}[q_{1}]\right)  _{x}%
\]
with the differential functions $q_{2}[q_{1}]$ and $q_{3}[q_{1}]$ given as
above. After substitution $u=-q_{1,xx}$ the second equation turns into the
well known Harry Dym equation while the first one becomes \ the second member
of the hierarchy. If we want to produce a next member of this hierarchy we
have to take $s=3$. According to the general remarks in the previous section
this new system will appear as the first system in our sequence of systems.

Let us now consider two-field systems: $N=\sigma+\alpha=2$. As before, we have
now two choices: $(\sigma,\alpha)=(1,1)$ and $(\sigma,\alpha)=(2,0)$. We start
with $(\sigma,\alpha)=(2,0)$. We have now $n=s+2\sigma-1=5$ and thus we
consider the Lagrangian $\mathcal{L}^{n,n-\sigma,-n}=\mathcal{L}^{5,3,-5}$.
The associated Euler-Lagrange equations (\ref{ELn}) can be written as%
\begin{align*}
q_{5}^{2}\left(  q_{1,x}^{2}+2q_{1,xx}q_{1}-2q_{2,xx}\right)  -2  &  =0\text{,
\ \ }2q_{4}-q_{5}^{3}q_{1,xx}=0\text{, }\\
4q_{3}q_{5}-q_{5,x}^{2}q_{5}^{2}-6q_{4}^{2}+2q_{5}^{3}q_{5,xx}  &  =0
\end{align*}
and they can be solved to%
\begin{align*}
q_{5}  &  =f_{5}^{5}[q_{1},q_{2}]=2w^{-1/2}\\
q_{4}  &  =f_{4}^{5}[q_{1},q_{2}]=4q_{1,xx}w^{-3/2}\\
q_{3}  &  =f_{3}^{5}[q_{1},q_{2}]=\left(  -\tfrac{5}{2}w_{x}^{2}%
+12q_{1,xx}^{2}w+2ww_{xx}\right)  w^{-7/2}%
\end{align*}
where $w=2q_{1,x}^{2}+4q_{1,xx}q_{1}-4q_{2,xx}$. Substituting it into
(\ref{Kna}) we arrive at the following two commuting two-component systems:%

\begin{align*}
q_{1,t_{3}}  &  =\left(  f_{3}^{5}[q_{1},q_{2}]\right)  _{x}\\
q_{2,t_{3}}  &  =q_{1}\left(  f_{3}^{5}[q_{1},q_{2}]\right)  _{x}-\left(
f_{3}^{5}[q_{1},q_{2}]\right)  q_{1,x}+\left(  f_{4}^{5}[q_{1},q_{2}]\right)
_{x}%
\end{align*}
and%
\begin{align}
q_{1,t_{4}}  &  =\left(  f_{4}^{5}[q_{1},q_{2}]\right)  _{x}\label{cHD}\\
q_{2,t_{4}}  &  =q_{1}\left(  f_{4}^{5}[q_{1},q_{2}]\right)  _{x}-\left(
f_{4}^{5}[q_{1},q_{2}]\right)  q_{1,x}+\left(  f_{5}^{5}[q_{1},q_{2}]\right)
_{x}\nonumber
\end{align}
The system (\ref{cHD}) can be written more explicitly as%
\begin{align*}
q_{1,t_{4}}  &  =2\left(  2q_{1,xxx}w-3q_{1,xx}w_{x}\right)  w^{-5/2}\\
q_{2,t_{4}}  &  =\left(  4q_{1}q_{1,xxx}w-6q_{1}q_{1,xx}w_{x}-4q_{1,x}%
q_{1,xx}w-ww_{x}\right)  w^{-5/2}.
\end{align*}
Finally, let us consider the case $(\sigma,\alpha)=(1,1)$. Again, we have
$n=5$, but this time we consider the Lagrangian $\mathcal{L}^{n,n-\sigma
-\alpha,-n+\alpha}=\mathcal{L}^{5,3,-4}$. Its first $n-\sigma-\alpha=3$
Euler-Lagrange equations
\begin{align*}
0  &  =-q_{5}^{2}q_{1,xx}-1\\
0  &  =4q_{4}+2q_{5}^{2}q_{5,xx}-q_{5}q_{5,x}^{2}\\
0  &  =2q_{3}q_{5}-q_{5}^{2}q_{4,x}q_{5,x}+q_{4}q_{5}q_{5,x}^{2}-3q_{4}%
^{2}+q_{5}^{3}q_{4,xx}-q_{5}^{2}q_{4}q_{5,xx}%
\end{align*}
yield the following elimination equations:%

\begin{align}
q_{5}  &  =f_{5}^{5}[q_{1}]=\left(  -q_{1,xx}\right)  ^{-1/2}\nonumber\\
q_{4}  &  =f_{4}^{5}[q_{1}]=-\tfrac{1}{2^{4}}\left(  5q_{1,xxx}^{2}%
-4q_{1,xx}q_{1,xxxx}\right)  \left(  -q_{1,xx}\right)  ^{-7/2} \label{niewiem}%
\\
q_{3}  &  =f_{3}^{5}[q_{1}]=\tfrac{1}{2^{9}}P[q_{1}]\left(  -q_{1,xx}\right)
^{-13/2}\nonumber
\end{align}
where $P[q_{1}]$ is some complicated differential polynomial of $q_{1}$
(homogeneous of degree $4$ and of order $6$) with integer coefficients.
Substituting (\ref{niewiem}) into the Killing systems (\ref{Kna}) we arrive at
the following two commuting two-component flows:%

\begin{align*}
q_{1,t_{3}}  &  =\left(  f_{3}^{5}[q_{1}]\right)  _{x}\\
q_{2,t_{3}}  &  =q_{1}\left(  f_{3}^{5}[q_{1}]\right)  _{x}-f_{3}^{5}%
[q_{1}]q_{1,x}+\left(  f_{4}^{5}[q_{1}]\right)  _{x}%
\end{align*}
and%
\begin{align*}
q_{1,t_{4}}  &  =\left(  f_{4}^{5}[q_{1}]\right)  _{x}\\
q_{2,t_{4}}  &  =q_{1}\left(  f_{4}^{5}[q_{1}]\right)  _{x}-f_{4}^{5}%
[q_{1}]q_{1,x}+\left(  f_{5}^{5}[q_{1}]\right)  _{x}%
\end{align*}
The last vector field can be written more explicitly as%

\begin{align*}
q_{1,t_{4}}  &  =\tfrac{1}{2^{5}}\left(  -40ww_{x}w_{xx}+35w_{x}^{3}%
+8w^{2}w_{xxx}\right)  w^{-9/2}\\
q_{2,t_{4}}  &  =\tfrac{1}{2^{5}}\left(  10q_{1,x}ww_{x}^{2}-8q_{1,x}%
w^{2}w_{xx}-40q_{1}ww_{x}w_{xx}+35q_{1}w_{x}^{3}+8q_{1}w^{2}w_{xxx}\right)
w^{-9/2}%
\end{align*}
where $w=-q_{1,xx}$. Let us notice that in this case we obtain a hierarchy of
systems such that every system is driven by its first equation which is a
consecutive equation of Harry Dym hierarchy. One can see that, contrary to the
positive case, if $\alpha>0$ the obtained systems are always driven by its
first $\sigma$ components that coincide with the corresponding systems from
$\alpha=0$ hierarchy.

\section{Conclusions}

In this paper we developed a method of unified constructing of St\"{a}ckel
systems and soliton hierarchies from the same common denominator in the form
of separation relations (\ref{2.7}). We developed our theory starting from
separation relations generated by separation curves of the form
\begin{equation}
H_{1}\lambda^{\beta_{1}}+...+H_{n}\lambda^{\beta_{n}}=\frac{1}{2}\lambda
^{m}\mu^{2}+\lambda^{k},\ \ \ \ \ \ \ \ \beta_{i},n\in\mathbf{N},\text{
\ }m,k\in\mathbf{Z}. \label{c1}%
\end{equation}
We performed a detailed, systematic construction of soliton hierarchies for
the Benenti class of separation relations, given by the particular form of
(\ref{c1}), namely
\[
H_{1}\lambda^{n-1}+H_{2}\lambda^{n-2}+\cdots+H_{n}=\frac{1}{2}\lambda^{m}%
\mu^{2}+\lambda^{k}.
\]
The results we obtained are hopefully only a first step of a new research
program. The next step of this program would be finding out a way for
systematic constructing of other soliton hierarchies from different classes of
separation curves (\ref{c1}), when the sequence $(\beta_{1},...,\beta_{n})$
differs from $(n-1,...,0)$. The next - nontrivial - step would be to extend
the theory to the case of polynomial separation curves (\ref{2.8}) with
$(\alpha_{1},...,\alpha_{n})\neq(0,...,0).$ We expect by presented procedure
to generate not only the majority of known soliton systems but also to
construct in a systematic way a vast number of new integrable hierarchies. One
should also investigate the possibility of "prolongation" of standard
integrable structures of separable systems (such as integrals of motion,
bi-Hamiltonian structure) onto the corresponding evolutionary hierarchies of PDE's.

\section{Appendix A}

The involutivity of $H_{1}^{(m,k)}$ and $H_{r}^{(m,k)}$ (\ref{Ham}) leads to
the following relations imposed on $g_{kk}^{(m)}(\lambda),$ $v_{r}^{k}%
(\lambda),\ V_{1}^{(k)}(\lambda)$ and $V_{r}^{(k)}(\lambda)$ \cite{ferap}:%
\begin{equation}
\frac{\partial v_{r}^{i}}{\partial\lambda_{i}}=0,\ \ \ i=1,...,n,
\tag{A1}\label{A1}%
\end{equation}%
\begin{equation}
\frac{\partial}{\partial\lambda_{i}}\ln g_{kk}^{(m)}=\frac{\frac{\partial
v_{r}^{k}}{\partial\lambda_{i}}}{v_{r}^{k}-v_{r}^{i}},\text{ }i\neq k,\text{
\ all }m,\text{ }r\text{ } \tag{A2}\label{A2}%
\end{equation}%
\begin{equation}
\frac{\partial V_{r}^{(k)}}{\partial\lambda_{i}}=v_{r}^{i}\frac{\partial
V_{1}^{(k)}}{\partial\lambda_{i}}\ \ \ \text{for all }i\text{, }r\text{ and
}k\text{.} \tag{A3}\label{A3}%
\end{equation}
We will prove here the relation (\ref{wzorek}), i.e.
\begin{equation}%
{\textstyle\sum_{i=1}^{n}}
E_{i}\left(  \mathcal{L}^{(n,m,k)}\right)  \lambda_{i,t_{r}}-%
{\textstyle\sum_{i=1}^{n}}
E_{i}\left(  \mathcal{L}_{r}^{(n,m,k)}\right)  \lambda_{i,x}=0. \tag{A4}%
\label{wzorekprim}%
\end{equation}
First, let us consider the geodesic case. Due to (\ref{Fer}) we have
$\lambda_{i,t_{r}}=v_{r}^{i}\lambda_{i,x}$ where $v_{r}^{i}$ are eigenvalues
of $K_{r}$ (see (\ref{Kdiag})). For the geodesic Hamiltonians
\[
\mathcal{L}^{(n,m,0)}=\frac{1}{2}\sum\limits_{i=1}^{n}g_{ii}^{(m)}%
\lambda_{i,x}^{2}\text{ \ and \ }\mathcal{L}_{r}^{(n,m,0)}=\frac{1}{2}%
\sum\limits_{i=1}^{n}g_{ii}^{(m)}v_{r}^{i}\lambda_{i,x}^{2}%
\]
so that in this case the left hand side of (\ref{wzorekprim}) attains the form%
\begin{align*}
&
{\textstyle\sum_{i=1}^{n}}
E_{i}\left(  \mathcal{L}^{(n,m,0)}\right)  \lambda_{i,t_{r}}-%
{\textstyle\sum_{i=1}^{n}}
E_{i}\left(  \mathcal{L}_{r}^{(n,m,0)}\right)  \lambda_{i,x}\\
&  =\left(  \frac{1}{2}\sum\limits_{i,k=1}^{n}\frac{\partial g_{kk}^{(m)}%
}{\partial\lambda_{i}}\left(  \lambda_{k,x}\right)  ^{2}v_{r}^{i}\lambda
_{i,x}-\sum\limits_{i=1}^{n}\frac{d}{dx}\left(  g_{ii}^{(m)}\lambda
_{i,x}\right)  v_{r}^{i}\lambda_{i}^{x}\right) \\
&  -\left(  \frac{1}{2}\sum\limits_{i,k=1}^{n}\frac{\partial g_{kk}^{(m)}%
}{\partial\lambda_{i}}\left(  \lambda_{k,x}\right)  ^{2}v_{r}^{k}\lambda
_{i,x}+\frac{1}{2}\sum\limits_{i,k=1}^{n}g_{kk}^{(m)}\frac{\partial v_{r}^{k}%
}{\partial\lambda_{i}}\left(  \lambda_{k,x}\right)  ^{2}\lambda_{i,x}\right.
\\
&  \left.  -\sum\limits_{i=1}^{n}\frac{d}{dx}\left(  g_{ii}^{(m)}\lambda
_{i,x}\right)  v_{r}^{i}\lambda_{i.x}-\sum\limits_{i=1}^{n}g_{ii}^{(m)}%
\lambda_{i,x}^{2}\frac{dv_{r}^{i}}{dx}\right) \\
&  =\frac{1}{2}\sum\limits_{i,k=1}^{n}\frac{\partial g_{kk}^{(m)}}%
{\partial\lambda_{i}}\left(  \lambda_{k,x}\right)  ^{2}\left(  v_{r}^{i}%
-v_{r}^{k}\right)  \lambda_{i,x}-\left(  \frac{1}{2}\sum\limits_{i,k=1}%
^{n}g_{kk}^{(m)}\frac{\partial v_{r}^{k}}{\partial\lambda_{i}}\left(
\lambda_{k,x}\right)  ^{2}\lambda_{i,x}-\sum\limits_{i=1}^{n}g_{ii}%
^{(m)}\lambda_{i,x}^{2}\frac{dv_{r}^{i}}{dx}\right)  =0
\end{align*}
since expression in the last parenthesis equals to%
\begin{align*}
&  \frac{1}{2}\sum\limits_{i,k=1}^{n}g_{kk}^{(m)}\frac{\partial v_{r}^{k}%
}{\partial\lambda_{i}}\left(  \lambda_{k,x}\right)  ^{2}\lambda_{i,x}%
-\sum\limits_{i,k=1}^{n}g_{ii}^{(m)}\lambda_{i,x}^{2}\frac{\partial v_{r}^{i}%
}{\partial\lambda_{k}}\lambda_{k,x}\\
&  =-\frac{1}{2}\sum\limits_{i,k=1}^{n}g_{kk}^{(m)}\frac{\partial v_{r}^{k}%
}{\partial\lambda_{i}}\left(  \lambda_{k,x}\right)  ^{2}\lambda_{i,x}%
=+\frac{1}{2}\sum\limits_{i,k=1}^{n}\frac{\partial g_{kk}^{(m)}}%
{\partial\lambda_{i}}\left(  \lambda_{k,x}\right)  ^{2}\left(  v_{r}^{i}%
-v_{r}^{k}\right)  \lambda_{i,x},
\end{align*}
where the last equality follows from formula (\ref{A2}) which can be written
in equivalent form%
\[
g_{kk}^{(m)}\frac{\partial v_{r}^{k}}{\partial\lambda_{i}}=\frac{\partial
g_{kk}^{(m)}}{\partial\lambda_{i}}\left(  v_{r}^{k}-v_{r}^{i}\right)
\]
(that is in fact also valid for $k=i$). Thus, the statement has been proved
for geodesic densities. For the potential parts:%
\[%
{\textstyle\sum_{i=1}^{n}}
E_{i}\left(  V_{1}^{(k)}\right)  \lambda_{i,t_{r}}-%
{\textstyle\sum_{i=1}^{n}}
E_{i}\left(  V_{r}^{(k)}\right)  \lambda_{i,x}=\sum\limits_{i=1}^{n}\left(
\frac{\partial V_{1}^{(k)}}{\partial\lambda_{i}}v_{r}^{i}-\frac{\partial
V_{r}^{(k)}}{\partial\lambda_{i}}\right)  \lambda_{i,x}=0
\]
due to (\ref{A3}). This concludes the proof.

\section{Appendix B}

We will prove here Theorem \ref{lemma5}. The relation (\ref{wtyl}) is a
consequence of (\ref{8a}) and (\ref{gmij}) of Lemma \ref{1}. Indeed, by
(\ref{gmij})
\[
E_{l}(\mathcal{L}^{n,m,k})=\frac{1}{2}\sum_{i,j=1}^{n}\frac{\partial
V_{1}^{(2n-m-i-j)}}{\partial q_{l}}\left(  q_{i}\right)  _{x}\left(
q_{j}\right)  _{x}-\frac{\partial V_{1}^{(k)}}{\partial q_{l}}-\frac{d}%
{dx}\left(  \sum_{i=1}^{n}V_{1}^{(2n-m-i-l)}\left(  q_{i}\right)  _{x}\right)
,
\]
and%
\begin{align*}
E_{l-\alpha}(\mathcal{L}^{n,m+\alpha,k-\alpha})  &  =\frac{1}{2}\sum
_{i,j=1}^{n}\frac{\partial V_{1}^{(2n-m-\alpha-i-j)}}{\partial q_{l-\alpha}%
}\left(  q_{i}\right)  _{x}\left(  q_{j}\right)  _{x}-\frac{\partial
V_{1}^{(k-\alpha)}}{\partial q_{l-\alpha}}\\
&  -\frac{d}{dx}\left(  \sum_{i=1}^{n}V_{1}^{(2n-m-\alpha-i-l+\alpha)}\left(
q_{i}\right)  _{x}\right)  \overset{\text{lemma}\ \text{\ref{1}}}{=}%
\ E_{l}(\mathcal{L}^{n,m,k}).
\end{align*}
The relation (\ref{wprzod}) is just a rewritten form of (\ref{wtly}).

Since in what follows we will compare separable potentials with different $n$,
in the rest of the proof we will use temporary extended notation for
potentials in the form $V_{1}^{n,(k)}.$ From (\ref{first}) and (\ref{16a}) it
follows that
\begin{equation}
V_{1}^{n,(n+k)}=V_{1}^{n+1,(n+1+k)},\ \ \ \ \ k=-n,...,n-1. \tag{B1}\label{20}%
\end{equation}
We prove now (\ref{19a}). Using the relation (\ref{8a}) for $r=1$ we obtain
\begin{align}
E_{l}\left(  \mathcal{L}^{n,0,0}\right)   &  =\frac{1}{2}\sum_{i,j=1}^{n}%
\frac{\partial V_{1}^{n,(2n-i-j)}}{\partial q_{l}}\left(  q_{i}\right)
_{x}\left(  q_{j}\right)  _{x}-\frac{d}{dx}\left(  \sum_{i=1}^{n}%
V_{1}^{n,(2n-i-l)}\left(  q_{i}\right)  _{x}\right) \nonumber\\
&  =\frac{1}{2}\sum_{i,j=1}^{n}\frac{\partial V_{1}^{n,(2n-i-j-l+1)}}{\partial
q_{1}}\left(  q_{i}\right)  _{x}\left(  q_{j}\right)  _{x}-\frac{d}{dx}\left(
\sum_{i=1}^{n}V_{1}^{n,(2n-i-l)}\left(  q_{i}\right)  _{x}\right)
\tag{B2}\label{16c}%
\end{align}
and in a similar way we have
\[
E_{l+1}\left(  \mathcal{L}^{n+1,0,0}\right)  =\frac{1}{2}\sum_{i,j=1}%
^{n+1}\frac{\partial V_{1}^{n+1,(2n+2-i-j-l)}}{\partial q_{1}}\left(
q_{i}\right)  _{x}\left(  q_{j}\right)  _{x}-\frac{d}{dx}\left(  \sum
_{i=1}^{n+1}V_{1}^{n+1,(2n-i-l+1)}\left(  q_{i}\right)  _{x}\right)
\]%
\[
\overset{(\ast)}{=}\frac{1}{2}\sum_{i,j=1}^{n}\frac{\partial V_{1}%
^{n+1,(2n-i-j-l+2)}}{\partial q_{1}}\left(  q_{i}\right)  _{x}\left(
q_{j}\right)  _{x}-\frac{d}{dx}\left(  \sum_{i=1}^{n}V_{1}^{n+1,(2n-i-l+1)}%
\left(  q_{i}\right)  _{x}\right)
\]%
\[
\overset{(\ref{20})}{=}\frac{1}{2}\sum_{i,j=1}^{n}\frac{\partial
V_{1}^{n,(2n-i-j-l+1)}}{\partial q_{1}}\left(  q_{i}\right)  _{x}\left(
q_{j}\right)  _{x}-\frac{d}{dx}\left(  \sum_{i=1}^{n}V_{1}^{n,(2n-i-l)}\left(
q_{i}\right)  _{x}\right)  .
\]
The equality $(\ast)$ is due to he fact that $V_{1}^{n+1,(2n-i-l+1)}=0$ for
$i=n+1$ and similarly $V_{1}^{n+1,(2n-i-j-l+2)}$ does not depend on $q_{1}$
for $i=n+1$ or $j=n+1$ (this follows from (\ref{19a}) and (\ref{16aa})). Thus
\begin{equation}
E_{l}\left(  \mathcal{L}^{n,0,0}\right)  =E_{l+1}\left(  \mathcal{L}%
^{n+1,0,0}\right)  ,\ \ \ l=1,...,n. \tag{B3}\label{a}%
\end{equation}
Moreover, from (\ref{6}) it follows that
\[
V_{1}^{n,(2n+\sigma)}=-q_{1}V_{1}^{n,(2n+\sigma-1)}-...-q_{n}V_{1}%
^{n,(n+\sigma)},
\]
hence, for $l>\sigma$
\begin{align*}
\frac{\partial V_{1}^{n,(2n+\sigma)}}{\partial q_{l}}  &  =-q_{1}%
\frac{\partial V_{1}^{n,(2n+\sigma-1)}}{\partial q_{l}}-...-q_{n}%
\frac{\partial V_{1}^{n,(n+\sigma)}}{\partial q_{l}}-V_{1}^{n,(2n+\sigma-l)}\\
& \\
&  =-q_{1}\frac{\partial V_{1}^{n,(2n+\sigma-l)}}{\partial q_{1}}%
-...-q_{n}\frac{\partial V_{1}^{n,(n+\sigma-l+1)}}{\partial q_{1}}%
-V_{1}^{n,(2n+\sigma-l)}.
\end{align*}
On the other hand we have
\[
V_{1}^{n+1,(2n+\sigma+2)}=-q_{1}V_{1}^{n+1,(2n+\sigma+1)}-...-q_{n}%
V_{1}^{n+1,(n+\sigma+2)}-q_{n+1}V_{1}^{n+1,(n+\sigma+1)},
\]
hence, for $l>\sigma$ and according to (\ref{20}) and (\ref{16a})
\begin{align}
\frac{\partial V_{1}^{n+1,(2n+\sigma+2)}}{\partial q_{l+1}}  &  =-q_{1}%
\frac{\partial V_{1}^{n+1,(2n+\sigma+1)}}{\partial q_{l+1}}-...-q_{n}%
\frac{\partial V_{1}^{n+1,(n+\sigma+2)}}{\partial q_{l+1}}-V_{1}%
^{n+1,(2n+\sigma-l+1)}\nonumber\\
& \nonumber\\
&  =-q_{1}\frac{\partial V_{1}^{n+1,(2n+\sigma+1-l)}}{\partial q_{1}%
}-...-q_{n}\frac{\partial V_{1}^{n+1,(n+\sigma+2-l)}}{\partial q_{1}}%
-V_{1}^{n+1,(2n+\sigma-l+1)}\nonumber\\
& \nonumber\\
&  =-q_{1}\frac{\partial V_{1}^{n,(2n+\sigma-l)}}{\partial q_{1}}%
-...-q_{n}\frac{\partial V_{1}^{n,(n+\sigma+1-l)}}{\partial q_{1}}%
-V_{1}^{n,(2n+\sigma-l)}\nonumber\\
& \nonumber\\
&  =\frac{\partial V_{1}^{n,(2n+\sigma)}}{\partial q_{l}}, \tag{B4}\label{b}%
\end{align}
and from (\ref{8a}) it follows that
\[
\frac{\partial V_{1}^{n,(n+s)}}{\partial q_{l}}=\frac{\partial V_{1}%
^{n+1,(n+s+2)}}{\partial q_{l+1}},\ \ \ \ \ 0\leq s-l+1\leq n.
\]
So, from (\ref{a}) and (\ref{b}) for $\sigma<l\leq n$ equation (\ref{19a}) is fulfilled.

Now, we pass to the proof of relations (\ref{19a1}). First, we have
\begin{align}
E_{l}\left(  \mathcal{L}^{n,n-\sigma,0}\right)   &  =\frac{1}{2}\sum
_{i,j=1}^{n}\frac{\partial V_{1}^{n,(n+\sigma-i-j)}}{\partial q_{l}}\left(
q_{i}\right)  _{x}\left(  q_{j}\right)  _{x}-\frac{d}{dx}\left(  \sum
_{i=1}^{n}V_{1}^{n,(n+\sigma-i-l)}\left(  q_{i}\right)  _{x}\right)
\nonumber\\
&  =\frac{1}{2}\sum_{i,j=1}^{n}\frac{\partial V_{1}^{n,(n+\sigma-i-j-l+1)}%
}{\partial q_{1}}\left(  q_{i}\right)  _{x}\left(  q_{j}\right)  _{x}-\frac
{d}{dx}\left(  \sum_{i=1}^{n}V_{1}^{n,(n+\sigma-i-l)}\left(  q_{i}\right)
_{x}\right)  . \tag{B5}\label{do1}%
\end{align}
On the other hand
\[
E_{l}\left(  \mathcal{L}^{n+1,n+1-\sigma,0}\right)  =\frac{1}{2}\sum
_{i,j=1}^{n+1}\frac{\partial V_{1}^{n+1,(n+\sigma-i-j-l+2)}}{\partial q_{1}%
}\left(  q_{i}\right)  _{x}\left(  q_{j}\right)  _{x}-\frac{d}{dx}\left(
\sum_{i=1}^{n+1}V_{1}^{n+1,(n+\sigma-i-l+1)}\left(  q_{i}\right)  _{x}\right)
.
\]
By (\ref{19a}) and (\ref{16aa}), for $l\leq\sigma$ the last term in both sums
does not contribute. Moreover, according to (\ref{20}) and the fact that
$V_{1}^{n,(2n)}-q_{n+1}=V_{1}^{n+1,(2n+1)},$ we have
\[
\frac{\partial V_{1}^{n,(n+k)}}{\partial q_{1}}=\frac{\partial V_{1}%
^{n+1,(n+k+1)}}{\partial q_{1}},\ \ \ \ \ k=-n,...,n,
\]
hence
\begin{align*}
E_{l}\left(  \mathcal{L}^{n+1,n+1-\sigma,0}\right)   &  =\frac{1}{2}%
\sum_{i,j=1}^{n}\frac{\partial V_{1}^{n,(n+\sigma-i-j-l+1)}}{\partial q_{1}%
}\left(  q_{i}\right)  _{x}\left(  q_{j}\right)  _{x}-\frac{d}{dx}\left(
\sum_{i=1}^{n}V_{1}^{n,(n+\sigma-i-l)}\left(  q_{i}\right)  _{x}\right) \\
&  =E_{l}\left(  \mathcal{L}^{n,n-\sigma,0}\right)  .
\end{align*}
Finally, we prove the relation (\ref{19a2}). From the negative recursion
(\ref{rekdown}), we have
\begin{equation}
\left.  V_{1}^{n,(-k)}(q_{n-k+1},...,q_{n})\right\vert _{q_{i}\rightarrow
q_{i+1},\ i=n-k+1,...,n}=V_{1}^{n+1,(-k)}(q_{n-k+2},\ldots,q_{n+1}%
),\ \ k=1,...,n. \tag{B6}\label{B4}%
\end{equation}
Then
\begin{align}
E_{\sigma+l}\left(  \mathcal{L}^{n,n-\sigma,0}\right)   &  =\frac{1}{2}%
\sum_{i,j=1}^{n}\frac{\partial V_{1}^{n,(n-i-j+1-\sigma)}}{\partial
q_{l+\sigma}}\left(  q_{i}\right)  _{x}\left(  q_{j}\right)  _{x}-\frac{d}%
{dx}\left(  \sum_{i=1}^{n}V_{1}^{n,(n-i-l)}\left(  q_{i}\right)  _{x}\right)
\nonumber\\
&  =\frac{1}{2}\sum_{i,j=1}^{n}\frac{\partial V_{1}^{n,(2n-i-j-l+1)}}{\partial
q_{n}}\left(  q_{i}\right)  _{x}\left(  q_{j}\right)  _{x}-\frac{d}{dx}\left(
\sum_{i=1}^{n}V_{1}^{n,(n-i-l)}\left(  q_{i}\right)  _{x}\right)
\tag{B7}\label{do2}%
\end{align}
and
\begin{align*}
E_{\sigma+l}\left(  \mathcal{L}^{n+1,n+1-\sigma,0}\right)   &  =\frac{1}%
{2}\sum_{i,j=1}^{n+1}\frac{\partial V_{1}^{n+1,(n-i-j+2-\sigma)}}{\partial
q_{l+\sigma}}\left(  q_{i}\right)  _{x}\left(  q_{j}\right)  _{x}-\frac{d}%
{dx}\left(  \sum_{i=1}^{n+1}V_{1}^{n+1,(n-i-l+1)}\left(  q_{i}\right)
_{x}\right) \\
&  \frac{1}{2}\sum_{i,j=1}^{n+1}\frac{\partial V_{1}^{n+1,(2n-i-j-l+3)}%
}{\partial q_{n+1}}\left(  q_{i}\right)  _{x}\left(  q_{j}\right)  _{x}%
-\frac{d}{dx}\left(  \sum_{i=1}^{n+1}V_{1}^{n+1,(n-i-l+1)}\left(
q_{i}\right)  _{x}\right) \\
&  \frac{1}{2}\sum_{i,j=0}^{n}\frac{\partial V_{1}^{n+1,(2n-i-j-l+1)}%
}{\partial q_{n+1}}\left(  q_{i+1}\right)  _{x}\left(  q_{j+1}\right)
_{x}-\frac{d}{dx}\left(  \sum_{i=0}^{n}V_{1}^{n+1,(n-i-l)}\left(
q_{i+1}\right)  _{x}\right)  .
\end{align*}
As for $i,j=0$ there is no contribution to the sum, so according to (\ref{B4})
we have
\begin{equation}
E_{\sigma+l}(\mathcal{L}^{n,n-\sigma,0})_{|q_{j}\rightarrow q_{j+1}}%
=E_{\sigma+l}\left(  \mathcal{L}^{n+1,n+1-\sigma,0}\right)
,\ \ l=1,...,n-\sigma,\ j=1,...,n. \tag{B8}\label{B5}%
\end{equation}
From (\ref{8a}) and (\ref{B4}) we have%
\begin{equation}
\left.  \frac{\partial V_{1}^{n,(-n)}}{\partial q_{i}}\right\vert
_{q_{j}\rightarrow q_{j+1,\ j=1,...,n}}=\frac{\partial V_{1}^{n+1,(-n)}%
}{\partial q_{i+1}}=\frac{\partial V_{1}^{n+1,(-n-1)}}{\partial q_{i}}\text{,
}i=1,...,n. \tag{B9}\label{B9}%
\end{equation}
that together with (\ref{B5}) proves the relation (\ref{19a2}).

\end{document}